\documentclass[11p]{article}
\usepackage{natbib}
\usepackage{amssymb,amsmath,amsthm,latexsym,amsfonts,amscd,dsfont,enumerate,color}
\usepackage{a4wide,graphicx,tikz}
\usetikzlibrary{arrows,shapes,shadows,fit,backgrounds}

\usepackage{booktabs}
\usepackage{array}
\usepackage[font=small,format=plain,labelfont=bf,up,textfont=normal,up,justification=justified,singlelinecheck=false]{caption}
\usepackage{multirow}
\usepackage{rotating}
\usepackage[english]{babel}

\usepackage{color}
\usepackage{ulem}

\tikzstyle{trader} = [circle, draw, top color=white, bottom color=blue!30, draw=blue!50!black!100, drop shadow, minimum height=4em]
\tikzstyle{bank} = [rectangle, draw, top color=white, bottom color=red!20, draw=red!50!black!100, drop shadow, rounded corners, minimum height=3em, text width=4em, text centered]
\tikzstyle{market} = [rectangle, draw, top color=white, bottom color=green!20, draw=green!50!black!100, drop shadow, rounded corners, minimum height=3em, text width=4em, text centered]
\tikzstyle{background} = [rectangle,fill=gray!10, inner sep=0.2cm, rounded corners=5mm]
\tikzstyle{line} = [draw, latex'-latex']
\tikzstyle{from} = [draw, latex'-]
\tikzstyle{to} = [draw, -latex']

\newcommand{\ind}[1]{\mathbf 1_{\{#1\}}}
\newcommand{\Exx}{\mathbb{E}}
\newcommand{\Ex}[2]{\mathbb{E}_{#1}\!\left[\,#2\,\right]}
\newcommand{\ExT}[3]{\mathbb{E}_{#1}^{#2}\!\left[\,#3\,\right]}

\newcommand{\rec}{R}
\newcommand{\lgd}{\mbox{L{\tiny GD}}}
\newcommand{\cva}{\mbox{C{\tiny VA}}}
\newcommand{\dva}{\mbox{D{\tiny VA}}}

\newtheorem{theorem}{Theorem}[section]
\newtheorem{definition}{Definition}[section]

\newtheorem{proposition}[theorem]{Proposition}
\newtheorem{remark}[theorem]{Remark}

\begin{document}

\title{{\large An updated version of this report will appear in the volume: 
{Veronesi, P. (Editor), \emph{Handbook in Fixed-Income Securities}, Wiley, 2014}} \\
{\bf \Large Nonlinear Valuation under Collateral, Credit Risk and Funding Costs: A Numerical Case Study Extending Black-Scholes}}
\author{
Damiano Brigo\thanks{Department of Mathematics, Imperial College, London, and Capco Institute}
\and
Qing Liu\thanks{Department of Mathematics, Imperial College, London}
\and
Andrea Pallavicini\thanks{Department of Mathematics, Imperial College, London, and Banca IMI, Milan}
%\and
%D. Perini\thanks{Mediobanca, Milan}
\and
David Sloth\thanks{Danske Bank, Copenhagen}
}
%\date{
%This version: \today.
%}

\maketitle

\begin{abstract}
We develop an arbitrage-free framework for consistent valuation of derivative trades with collateralization, counterparty credit gap risk, and funding costs, following the approach first proposed by Pallavicini and co-authors in 2011. Based on the risk-neutral pricing principle, we derive a general pricing equation where Credit, Debit, Liquidity and Funding Valuation Adjustments (CVA, DVA, LVA and FVA) are introduced by simply modifying the payout cash-flows of the deal. Funding costs and specific close-out procedures at default break the bilateral nature of the deal price and render the valuation problem a non-linear and recursive one. CVA and FVA are in general not really additive adjustments, and the risk for double counting is concrete. We introduce a new adjustment, called a Non-linearity Valuation Adjustment (NVA), to address double-counting. Our framework is based on real market rates, since the theoretical risk free rate disappears from our final equations. The framework addresses common market practices of ISDA governed deals without restrictive assumptions on collateral margin payments and close-out netting rules, and can be tailored also to CCP trading under initial and variation margins, as explained in detail in \cite{BrigoPallaviciniJFE}. In particular, we allow for asymmetric collateral and funding rates, replacement close-out and re-hypothecation. The valuation equation takes the form of a backward stochastic differential equation or semi-linear partial differential equation, and can be cast as a set of iterative equations that can be solved by least-squares Monte Carlo. We propose such a simulation algorithm in a case study involving a generalization of the benchmark model of Black and Scholes for option pricing. Our numerical results confirm that funding risk has a non-trivial impact on the deal price, and that double counting matters too. We conclude the article with an analysis of large scale implications of non-linearity of the pricing equations: non-separability of risks, aggregation dependence in valuation, and local pricing measures as opposed to universal ones. This prompts a debate and a comparison between the notions of price and value, and will impact the operational structure of banks. This paper is an evolution, in particular, of the work by \cite{pallavicini2011funding,pallavicini2012funding}, \cite{PallaviciniBrigo2013multicurva}, and \cite{Sloth2013}. 
\end{abstract}

\medskip

\noindent\textbf{AMS Classification Codes}: 62H20, 91B70 \newline

%\newline

\noindent \textbf{Keywords}: Credit Valuation Adjustment, Counterparty Credit Risk, Funding Valuation Adjustment, Funding Costs, Collateralization, Non-linearity Valuation Adjustment, Derivatives Pricing, CVA, DVA, LVA, FVA, NVA, Funding-DVA, semi-linear PDE, BSDE, Nonlinear Valuation, Nonlinear Feynman-Kac, Least-squares Monte Carlo, Black-Scholes

\pagestyle{myheadings} \markboth{}{{\footnotesize  Brigo, D., Liu, Q., Pallavicini, A., Sloth, D.  Nonlinear Valuation with Credit Risk, Collateral \& Funding Costs}}

\newpage
%\tableofcontents
%\newpage

\section{Introduction}

In the wake of the financial crisis in 2007-2008, dealers and financial institutions have been forced to rethink how they value and hedge contingent claims traded either in the over-the-counter (OTC) market or through central clearing (CCPs). OTC derivatives are bilateral financial contracts negotiated between two default-risky entities. Yet, prior to the crisis, institutions tended to ignore the credit risk of high-quality rated counterparties, but as recent history has shown this was a particularly dangerous assumption. Moreover, as banks became reluctant to lend to each other with the crisis rumbling through the Western economies, the spread between the rate on overnight indexed swaps (OISs) and the LIBOR rate blew up. 

To keep up with this sudden change of game, dealers today make a number of adjustments when they book OTC trades. The credit valuation adjustment (CVA) corrects the price for the expected costs to the dealer due to the possibility that the counterparty may default, while the so-called debt valuation adjustment (DVA) is a correction for the expected benefits to the dealer due to his own default risk. The latter adjustment has the controversial effect that the dealer can book a profit as his default risk increases and is very hard if not impossible to hedge. Finally, dealers often adjust the price for the costs of funding the trade. In the industry, this practice is known as a liquidity and funding valuation adjustment (LVA, FVA). When a derivatives desk executes a deal with a client, it backs the trade by hedging it with other dealers in the market and by posting/receiving collateral and by receiving/paying interest on the collateral posted. This involves borrowing or lending money and other assets. Classical derivatives pricing theory rests on the assumption that one can borrow and lend at a unique risk-free rate of interest, a theoretical risk-free rate that is proxied by a number of market rates. The seminal work of Black-Scholes-Merton showed that in this case an option on equity can be replicated by a portfolio of equity and risk-free debt over any short period of time. Prior to the crisis, this assumption may have been reasonable with banks funding their hedging strategies at LIBOR. However, with drastically increasing spreads emerging as the crisis took hold, it became apparent that LIBOR is contaminated by credit risk (besides fraud risk) and as such is an imperfect proxy of the risk-free rate.  While overnight rates have replaced LIBOR as proxies for the risk free rate, it would be preferable for a pricing framework not to feature theoretical rates in the final valuation equations.  

In this paper we develop an arbitrage-free framework for consistent valuation of collateralized as well as uncollateralized trades under counterparty credit risk, collateral margining and funding costs. The need to consistently account for the changed trading conditions in the valuation of derivatives is stressed by the sheer size of the OTC market.
Indeed, despite the crisis and the previously neglected risks, the size of derivatives markets remains staggering; the market value of outstanding OTC derivative contracts equaled \$24.7 trillion by the end of 2012 with a whopping \$632.6 trillion in notional value (\cite{bis2013}). Adopting the risk-neutral valuation principle, we derive a general pricing equation for a derivative deal where the new or previously neglected types of risks are included simply as modifications of the payout cash-flows. We address the current market practices in accordance with the guidelines of the International Swaps and Derivatives Association without assuming restrictive constraints on the collateral margining procedures and close-out netting rules. In particular, we allow for asymmetric collateral and funding rates as well as exogenously given liquidity policies and hedging strategies. We also discuss re-hypothecation of collateral guarantees. 

When dealing with funding costs, one may take a single deal (micro) or homogeneous (macro) cost view. In the micro view, funding costs are determined at deal level. This means that the trading desk may borrow funds at a different rate than at which it can invest funds, and the rates may vary across deals even in the same desk. 
In a slightly more aggregate cost view, average funding spreads are applied to all deals yet the spread on borrowing funds may still be different from that on lending. Finally, if we turn to the macro and symmetric view, funding costs of borrowing and lending are assumed the same and a common funding spread is applied across all deals. Clearly, the treasury department of a bank plays an active part in the micro approach and works as an operational center, while in the macro approach it takes more the role of a supporting function for the trading business.
In this work we stay as general as possible and adopt a micro cost view. Naturally, the macro view is just a special case of the micro view. This will be implicit in making the otherwise exogenously assigned funding rates a function of the specific deal value.  One should notice that the specific treasury model one adopts also impacts the presence of credit risk, and in particular of DVA, on the funding policy. This effect is occasionally referred to as DVA2, but we will not adopt such terminology here.  

The introduction of funding risk makes the pricing problem recursive and non-linear. The price of the deal depends on the trader's funding strategy in future paths, while to determine the future funding strategy we need to know the deal price itself in future paths.  This recursive structure was also discovered in the studies of \cite{pallavicini2011funding}, \cite{Crepey2011} and \cite{BurgardKjaer2011a}, yet the feature is neglected in the common approach of adding a funding spread to the discount curve.

Valuation under funding risk poses a significantly more complex and computationally demanding problem than standard CVA and DVA computations (except possibly CVA/DVA under replacement close-out), since it requires forward simulation and backward induction at the same time. In addition, FVA does not take the form of a simple additive term as appears to be commonly assumed by market participants. Funding and credit costs do not split up in a purely additive way. A consequence of this is that valuation becomes aggregation-dependent as portfolio prices do not add up. It is therefore difficult for banks to create CVA and FVA desks with separate and clear-cut responsibilities. This can be done at the expense of tolerating some degree of double counting. In this respect,  we introduce NVA, defined as a non-linearity valuation adjustment, measuring the degree of valuation error one has when removing non-linearities through symmetrization of borrowing and lending rates and through a risk-free close-out substituting a replacement close-out at default.  
More generally, non-linearity implies organizational challenges which we hint at in the conclusions.  

To thoroughly explore valuation under funding costs, we show how the general pricing equation can be cast as a set of iterative equations that can be conveniently solved by means of least-squares Monte Carlo (see, e.g., \cite{carriere1996valuation}, \cite{longstaff2001valuing}, \cite{tilley1993valuing}, and \cite{tsitsiklis2001regression}) and we propose an efficient simulation algorithm. Also, we derive a continuous-time approximation of the solution of the pricing equation as well as the associated partial differential equation (PDE), showing that the risk-free rate disappears. 

Despite its general market acceptance, the practice of  including an adjustment for funding costs has stirred quite some controversy among academics and practitioners (see the debate following \cite{HW2012a}).  At the center of this controversy is the issue that funding-contingent pricing becomes subjective due to asymmetric information. The particular funding policy chosen by the client is not (fully) known to the dealer, and vice versa. As a result, the price of the deal may be different to either of the two parties. Theoretically, this should mean that the parties would never close the deal. However, in reality, the dealer may not be able to recoup his full funding costs from the client, yet traders say that funding risk was the key factor driving bid-ask spreads wider during the crisis. 

The importance of the funding valuation adjustment in terms of impact has been stressed by J.P. Morgan's results in January 2014.
Michael Rapoport reports on January 14, 2014 in the Wall Street Journal:

\bigskip

{\it{"[...] So what is a “funding valuation adjustment,” and why did it cost J.P. Morgan Chase \$1.5 billion? The giant bank recorded a \$1.5 billion charge in its fourth-quarter earnings announced Tuesday because of the adjustment -- the result of a complex change in J.P. Morgan’s approach to valuing some of the derivatives on its books. J.P. Morgan was persuaded to make the “FVA” change by an “industry migration” toward such a move, the bank said in an investor presentation. A handful of other large banks, mostly in the U.K. and Europe, have already made a similar change."}}

\bigskip

In terms of available literature in this area, several studies have analyzed the various valuation adjustments separately, but few have tried to build a valuation approach that consistently takes collateralization, counterparty credit risk, and funding costs into account. Under unilateral default risk, ie. when only one party is defaultable, \cite{BrigoMasetti} consider valuation of derivatives with CVA, while particular applications of their approach are given in \cite{BrigoPallavicini2007}, \cite{BrigoChourdakis}, and \cite{BrigoMoriniTarenghi}, see \cite{BrigoMoriniPallavicini2012} for a summary. Bilateral default risk appears in \cite{BieleckiRutkowski2002}, \cite{BrigoCapponi2010}, \cite{BrigoPallaviciniPapatheodorou}, and  \cite{Gregory2009} who price both the CVA and DVA of a derivatives deal. The impact of collateralization on default risk has been investigated in \cite{Cherubini} and more recently in \cite{BrigoCapponiPallavicini} and \cite{BrigoCapponiPallaviciniPapatheodorou}. 

Assuming no default risk, \cite{Piterbarg2010} provides an initial analysis of collateralization and funding risk in a stylized Black-Scholes economy. Yet, the introduction of collateral in a world without default risk is questionable since its main purpose is to mitigate such risk. \cite{Fujii2010} analyze consequences of multi-currency features in collateral proceedings. \cite{MoriniPrampolini2011}, \cite{Fries2010} and \cite{Castagna2011} consider basic implications of funding in presence of default risk. However, these works focus only on simple financial products, such as zero-coupon bonds or loans, and do not offer the level of generality needed to include all the required features. Thus, a general framework for consistent valuation under the new risks is still missing. The most comprehensive attempts are those of \cite{BurgardKjaer2011a, BurgardKjaer2011b}, \cite{Crepey2011,Crepey2012a,Crepey2012b}, and \cite{pallavicini2011funding,pallavicini2012funding}. Nonetheless, as \cite{BurgardKjaer2011a, BurgardKjaer2011b} resort to a PDE approach, their results are constrained to low dimensions. Also, they neglect the hidden complexities of collateralization and mark-to-market discontinuities at default.
The approach of \cite{Crepey2011,Crepey2012a,Crepey2012b} is more general although it does not allow for credit instruments in the deal portfolio. 

We follow the work of \cite{pallavicini2011funding,pallavicini2012funding} and \cite{Sloth2013} and consider a general pricing framework for derivative deals that fully and consistently takes collateralization, counterparty credit risk, and funding risk into account. The framework is conceptually simple and intuitive in contrast to previous attempts. It is based on the celebrated risk-neutral valuation principle and the new risks are included simply by adjusting the payout cash-flows of the deal. We present  a numerical case study that extends the benchmark theory of Black-Scholes for equity options to credit gap risk (CVA/DVA after collateralization), collateral and funding costs. 
We find that the precise patterns of funding-inclusive values depend on a number of factors, including the asymmetry between borrowing and lending rates. We stress such inputs in order to analyze their impact on the funding-inclusive price. Our numerical results confirm that funding risk has a non-trivial impact on the deal price and that double counting can be relevant as well.

As we point out in the conclusions, the violation of the bilateral nature of valuation and the aggregation dependence lead to doubts about considering the funding-inclusive value as a price. Should this value be charged to a client? This prompts to the old distinction between price and value, and whether funding costs should be considered as a profitability/cost analysis tool rather than as an adjustment to be charged to a client. This is often reinforced by the fact that a bank client has no direct control on the funding policy of a bank. 

It is also worth pointing out that the theory we illustrate here on the Black--Scholes benchmark case has also been applied to pricing under multiple interest rate curves, see \cite{PallaviciniBrigo2013multicurva}, and to pricing under central clearing (CCPs) in presence of default risk of clearing members, close-out delays, and initial and variation margins, see \cite{BrigoPallaviciniJFE}. 

The paper is organized as follows. Section \ref{p4:sec:framework} describes the general pricing framework with collateralized credit, debit, liquidity and funding valuation adjustments. Section \ref{p4:sec:pricingEq} derives an iterative solution of the pricing equation as well as a continuous-time approximation. Section \ref{p4:sec:numerics} describes a least-square Monte Carlo algorithm and provides numerical results on deal positions in European call options on equity under the benchmark model of Black and Scholes. In addition, a non-linearity valuation adjustment, NVA, is introduced and computed. 
%Section \ref{p4:sec:extensions} extends the pricing framework to cases where the dealer hedges the trade using other derivatives. 
Finally, Section \ref{p4:sec:conclusion} concludes the paper.

\section{Collateralized Credit and Funding Valuation Adjustments}\label{p4:sec:framework}

In this section we develop a general risk-neutral valuation framework for OTC derivative deals. The section clarifies how the traditional pre-crisis derivative price is consistently adjusted to reflect the new market realities of collateralization, counterparty credit risk, and funding risk.
We refer to the two parties of a credit-risky deal as the investor or dealer ("I") on one side and the counterparty or client ("C") on the other. 

Fixing the time horizon $T\in\mathbb R_+$ of the deal, we define our risk-neutral pricing model on the probability space $(\Omega,\mathcal G,(\mathcal G_t)_{t\in[0,T]},\mathbb Q)$. $\mathbb{Q}$ is the risk-neutral probability measure. The filtration $(\mathcal G_t)_{t\in[0,T]}$ models the flow of information of the whole market, including credit, such that the default times of the investor $\tau_I$ and the counterparty $\tau_C$ are $\mathcal G$-stopping times. We adopt the notational convention that $\mathbb E_t$ is the risk-neutral expectation conditional on the information $\mathcal G_t$ while  $\mathbb E_{\tau_i}$ denotes the conditional risk-neutral expectation given the stopped filtration $\mathcal G_{\tau_i}$. Moreover, we exclude the possibility of simultaneous defaults for simplicity and define the time of the first default event among the two parties as the stopping time
\begin{equation*}
\tau \triangleq (\tau_I \wedge \tau_C).
\end{equation*}
In the sequel we adopt the view of the investor and consider the cash-flows and consequences of the deal from his perspective. In other words, when we price the deal we obtain the value of the position to the investor. As we will see, with funding risk this price will often not just be the value of the deal to the counterparty with opposite sign.

The gist of the valuation framework is conceptually simple and rests neatly on the classical finance disciplines of risk-neutral pricing and discounting cash-flows. When a dealer enters into a derivatives deal with a client, a number of cash-flows are exchanged, and just like valuation of any other financial claim, discounting these cash in- or outflows gives us a price of the deal. Post-crisis market practice distinguishes four different types of cash-flow streams occurring once a trading position has been entered: (i) Cash-flows coming directly from the derivatives contract such as payoffs, coupons, dividends, etc. We denote by $\pi(t,T)$ the sum of the discounted payoff happening over the time period $(t,T]$. This is where classical derivatives pricing would usually stop and the price of a derivative contract with maturity $T$ would be given by 
\begin{align}
{ V}_t =  \mathbb E_{t} & \left[ \, \pi(t,T) \right].\nonumber
\end{align}%
This price assumes no credit risk of the parties involved and no funding risk of the trade. However, present-day market practice requires the price to be adjusted by taking further cash-flow transactions into account: (ii) Cash-flows required by collateral margining. If the deal is collateralized, cash flows happen in order to maintain a collateral account that in the case of default will be used to cover any losses. $\gamma(t,T;C)$ is the sum of the discounted margining costs over the period $(t,T]$ with $C$ denoting the collateral account. (iii) Cash-flows exchanged once a default event has occurred. We let $\theta_\tau (C,\varepsilon)$ denote the on-default cash-flow with $\varepsilon$ being the residual value of the claim traded at default. Lastly, (iv) cash-flows required for funding the deal. We denote the sum of the discounted funding costs over the period $(t,T]$ by $\varphi(t,T;F)$ with $F$ being the cash account needed for funding the deal. Collecting the terms we obtain a consistent price $\bar V$ of a derivative deal taking into account counterparty credit risk, margining costs, and funding costs
\begin{align}
\label{p4:eq:fundingpreview}
{\bar V}_t(C,F) =  \mathbb E_{t} & \left[ \, \pi(t,T\wedge\tau) + \gamma(t,T\wedge\tau;C) + \varphi(t,T\wedge\tau;F) \right. \\ \nonumber
  & \left. + \ind{t<\tau<T} D(t,\tau) \theta_\tau(C,\varepsilon) \right],
\end{align}%
where $D(t,\tau)$ is the risk-free discount factor.

By using a risk-neutral valuation approach, we see that only the payout needs to be adjusted under counterparty credit and funding risk.
In the following paragraphs we expand the terms of \eqref{p4:eq:fundingpreview} and carefully discuss how to compute them.

\subsection{Trading under Collateralization and Close-out Netting}

The ISDA master agreement is the most commonly used framework for full and flexible documentation of OTC derivative transactions and is published by the International Swaps and Derivatives Association (\cite{ISDA2009}). Once agreed between two parties, the master agreement sets out standard terms that apply to all deals entered into between those parties. 
%Each time that a transaction is entered into, the terms of the master agreement do not need to be re-negotiated and apply automatically.
The ISDA master agreement lists two tools to mitigate counterparty credit risk: \textit{collateralization} and \textit{close-out netting}. Collateralization of a deal means that the party which is out-of-the-money is required to post collateral -- usually cash, government securities or highly rated bonds -- corresponding to the amount payable by that party in the case of a default event. The credit support annex (CSA) to the ISDA master agreement defines the rules under which the collateral is posted or transferred between counterparties. Close-out netting means that in the case of default all transactions with the counterparty under the ISDA master agreement are consolidated into a single net obligation which then forms the basis for any recovery settlements. 

\paragraph{Collateralization} 

Collateralization of a deal usually happens according to a margining procedure. Such a procedure involves that both parties post collateral amounts to or withdraw amounts from the collateral account $C$ according to their current exposure on pre-fixed dates $\{t_1,\ldots,t_n=T\}$ during the life of the deal, typically daily. Let $\alpha_i$ be the year fraction between $t_{i}$ and $t_{i+1}$. The terms of the margining procedure may, furthermore, include independent amounts, minimum transfer amounts, thresholds, etc., as described in \cite{BrigoCapponiPallaviciniPapatheodorou}. However, here we adopt a general description of the margining procedure that does not rely on the particular terms chosen by the parties.

We consider a collateral account $C$ held by the investor. Moreover, we assume that the investor is the collateral taker when $C_t>0$ and the collateral provider when $C_t<0$.
The CSA ensures that the collateral taker remunerates the account $C$ at an accrual rate. If the investor is collateral taker, he remunerates the collateral account by the accrual rate $c^+_t(T)$, while if he is the collateral provider, the counterparty remunerates the account at the rate $c^-_t(T)$\footnote{We stress the slight abuse of notation here: A plus and minus sign does not indicate that the rates are positive or negative parts of some other rate, but instead it tells which rate is used to accrue interest on the collateral according to the sign of the collateral account.}. The effective accrual collateral rate $\tilde c_t(T)$ is defined as
\begin{equation}
\label{p4:eq:ctilda}
{\tilde c}_t(T) \triangleq c^-_t(T) \ind{C_t<0} + c^+_t(T) \ind{C_t>0} \,,
\end{equation}
%and the corresponding zero-coupon bond is given by
%\[
%P^{\tilde c}_t(T) \triangleq \frac{1}{1+(T-t){\tilde c}_t(T)} \,.
%\]%
To understand the cash-flows originating from collateralization of the deal, let us consider the consequences of the margining procedure to the investor. At the first margin date, say $t_1$, the investor opens the account and posts collateral if he is out-of-the-money, i.e. if $C_{t_1}<0$, which means that the counterparty is the collateral taker. On each of the following margin dates $t_k$, the investor posts collateral according to his exposure as long as $C_{t_k}<0$. As collateral taker, the counterparty pays interest on the collateral at the accrual rate $c^-_{t_k}(t_{k+1})$ between the following margin dates $t_k$ and $t_{k+1}$. We assume that interest accrued on the collateral is saved into the account and thereby directly included in the margining procedure and the close-out. Finally, if $C_{t_n}<0$ on the last margin date $t_n$, the investor closes the collateral account given no default event has occurred in between. Similarly, for positive values of the collateral account, the investor is instead the collateral taker and the counterparty faces corresponding cash-flows at each margin date. If we sum up all the discounted margining cash-flows of the investor and the counterparty, we obtain 
\begin{equation}
\label{p4:eq:gamma}
\gamma(t,T\wedge\tau;C) \triangleq \sum_{k=1}^{n-1} \ind{t \leqslant t_k < (T\wedge\tau)} D(t,t_k) C_{t_k} \left( 1 - \frac{P_{t_k}(t_{k+1})}{P^{\tilde c}_{t_k}(t_{k+1})} \right) \,,
\end{equation}%
with the zero-coupon bond $P^{\tilde c}_t(T) \triangleq [1+(T-t){\tilde c}_t(T) ]^{-1}$. 
If we adopt a first order expansion  (for small $c$ and $r$)  we can approximate
\begin{equation}
\label{p4:eq:gamma2}
\gamma(t,T\wedge\tau;C) \approx \sum_{k=1}^{n-1} \ind{t \leqslant t_k < (T\wedge\tau)} D(t,t_k) C_{t_k} \alpha_k  \left(  r_{t_{k}}(t_{k+1}) - \tilde{c}_{t_{k}}(t_{k+1}) \right) \,,
\end{equation}%
where with a slight abuse of notation we call $\tilde{c}_t(T)$ and $r_t(T)$ the continuously (as opposed to simple) compounded interest rates associated with the bonds $P^{\tilde{c}}$ and $P$. This last expression clearly shows a cost of carry structure for collateral costs. If $C$ is positive to ``I", then ``I" is holding collateral and will have to pay (hence the minus sign) an interest $c^+$, while receiving the natural growth $r$ for cash, since we are in a risk neutral world. In the opposite case, if ``I" posts collateral, $C$ is negative to ``I" and ``I" receives interest $c^-$ while paying the risk free rate, as should happen when one shorts cash in a risk neutral world. 

A crucial role in collateral procedures is played by re-hypothecation. We discuss rehypothecation and its inherent liquidity risk in the following.

\paragraph{Rehypothecation} 

Often the CSA grants the collateral taker relatively unrestricted use of the collateral for his liquidity and trading needs until it is returned to the collateral provider. Effectively, the practice of rehypocthecation lowers the costs of remuneration of the provided collateral. However, while without rehypothecation the collateral provider can expect to get any excess collateral returned after honoring the amount payable on the deal, if  rehypothecation is allowed the collateral provider runs the risk of losing a fraction or all of the excess collateral in case of default on the collateral taker's part. 

We denote the recovery fraction on the rehypothecated collateral by $R'_I$ when the investor is the collateral taker and by $R'_C$ when the counterparty is the collateral taker. The general recovery fraction on the market value of the deal that the investor receives in the case of default of the counterparty is denoted by $R_C$, while $R_I$ is the recovery fraction received by the counterparty if the investor defaults. The collateral provider typically has precedence over other creditors of the defaulting party in getting back any excess capital, which means $R_I \leqslant R'_I \leqslant 1$ and $R_C \leqslant R'_C \leqslant 1$. 
If no rehypothecation is allowed and the collateral is kept safe in a segregated account, we have that $R'_I=R'_C=1$.

\paragraph{Close-out netting} 

In case of default all terminated transactions under the ISDA master agreement with a given counterparty are netted and consolidated into a single claim. This also includes any posted collateral to back the transactions. In this context the close-out amount plays a central role in calculating the on-default cash-flows. The close-out amount is the costs or losses that the surviving party incurs when replacing the terminated deal with an economic equivalent. Clearly, the size of these costs will depend on which party survives so we define the close-out amount as
\begin{equation}
\varepsilon_\tau \triangleq \ind{\tau=\tau_C<\tau_I} \varepsilon_{I,\tau} + \ind{\tau=\tau_I<\tau_C} \varepsilon_{C,\tau} \,,
\end{equation}
where $\varepsilon_{I,\tau}$ is the close-out amount on the counterparty's default priced at time $\tau$ by the investor and $\varepsilon_{C,\tau}$ is the close-out amount if the investor defaults. Recall that we always consider the deal from the investor's viewpoint in terms of the sign of the cash-flows involved. This means that if the close-out amount $\varepsilon_{I,\tau}$ as measured by the investor is positive, the investor is a creditor of the counterpaty, while if it is negative, the investor is a debtor of the counterparty. Analogously, if the close-out amount $\varepsilon_{C,\tau}$ to the counterparty but viewed from the investor is positive, the investor is a creditor of the counterparty, and if it is negative, the investor is a debtor to the counterparty.

We note that the ISDA documentation is, in fact, not very specific in terms of how to actually calculate the close-out amount. Since 2009 ISDA has allowed for the possibility to switch from a risk-free close-out rule to a replacement rule that includes the DVA of the surviving party in the recoverable amount. \cite{Parker2009} and \cite{Weeber2009} show how a wide range of values of the close-out amount can be produced within the terms of ISDA. We refer to \cite{BrigoCapponiPallaviciniPapatheodorou} and the references therein for further discussions on these issues. Here, we adopt the approach of \cite{BrigoCapponiPallaviciniPapatheodorou} listing the cash-flows of all the various scenarios that can occur if default happens. We will net the exposure against the pre-default value of the collateral $C_{\tau-}$ and treat any remaining collateral as an unsecured claim. 
If we aggregate all these cash-flows and the pre-default value of collateral account, we reach the following expression for the on-default cash-flow 
\begin{align}
\label{p4:eq:theta}
\theta_{\tau}(C,\varepsilon)
  \triangleq  & \quad \;\ind{\tau=\tau_C<\tau_I} \left(\varepsilon_{I,\tau} - \lgd_C (\varepsilon_{I,\tau}^+ - C_{\tau^-}^+)^+ - \lgd'_C (\varepsilon_{I,\tau}^- - C_{\tau^-}^-)^+ \right) \\\nonumber
 &  + \ind{\tau=\tau_I<\tau_C} \left(\varepsilon_{C,\tau} - \lgd_I (\varepsilon_{C,\tau}^- - C_{\tau^-}^-)^- -\lgd'_I (\varepsilon_{C,\tau}^+ - C_{\tau^-}^+)^- \right) \,.
\end{align}%
where we define the loss-given-default as $\lgd_C \triangleq 1-R_C$, and the collateral loss-given-default as $\lgd'_C \triangleq 1-R'_C$. If both parties agree on the exposure, namely $\varepsilon_{I,\tau} = \varepsilon_{C,\tau} =\varepsilon_{\tau}$, when we take the risk-neutral expectation in \eqref{p4:eq:fundingpreview}, we see that the price of the discounted on-default cash-flow, 
\begin{equation}\label{eq:cvadvacloseout}
\mathbb E_t[\ind{t<\tau<T}D(t,\tau)\theta_\tau(C,\varepsilon)]=\mathbb E_t[\ind{t<\tau<T}D(t,\tau)\,\varepsilon_\tau)]-\cva(t,T;C)+\dva(t,T;C), 
\end{equation}
 is the present value of the close-out amount reduced by the positive collateralized CVA and DVA terms
\begin{align*}
&\Pi_{\mbox{\footnotesize CVAcoll}}(s) =  \left(  \lgd_C (\varepsilon_{I,s}^+ - C_{s^-}^+)^+ + \lgd'_C (\varepsilon_{I,s}^- - C_{s-}^-)^+ \right)  \ge 0,\\
&\Pi_{\mbox{\footnotesize DVAcoll}}(s) = -\left( \lgd_I (\varepsilon_{C,s}^- - C_{s-}^-)^- + \lgd'_I (\varepsilon_{C,s}^+ - C_{s-}^+)^- \right)  \ge 0, 
\end{align*}
and
\begin{align}
\label{p4:eq:cvaterms}
\cva(t,T;C)  &\triangleq  \mathbb E_t \left[ \ind{\tau=\tau_C<T} D(t,\tau) \Pi_{\mbox{\footnotesize CVAcoll}}(\tau)\right] ,\nonumber\\
 \dva(t,T;C)  &\triangleq  \mathbb E_t \left[  \ind{\tau=\tau_I<T} D(t,\tau) \Pi_{\mbox{\footnotesize DVAcoll}}(\tau) \right]\,.
\end{align}
Also, observe that if rehypothecation of the collateral is not allowed, the terms multiplied by $\lgd'_C$ and $\lgd'_I$ drops out of the CVA and DVA calculations.

\subsection{Trading under Funding Risk}

The hedging strategy that perfectly replicates the no-arbitrage price of a derivative is formed by a position in cash and a position in a portfolio of hedging instruments. When we talk about a derivative deal's funding, we essentially mean the cash position that is required as part of the hedging strategy, and with funding costs we refer to the costs of maintaining this cash position. If we denote the cash account by $F$ and the risky-asset account by $H$, we get
\begin{equation*}
 {\bar V}_t = F_t + H_t \,.
\end{equation*}
In the classical Black-Scholes-Merton theory, the risky part $H$ of the hedge would be a delta position in the underlying stock, whereas the risk-less part $F$ would be a position in the risk-free bank account. %Thus, classical theory assumes that we can fund the trade --  i.e., borrow and invest cash -- at a unique risk-free rate.
If the deal is collateralized, the margining procedure is included in the deal definition insuring that funding of the collateral is automatically taken into account.
Moreover, if rehypothecation is allowed for the collateralized deal, the collateral taker can use the posted collateral as a funding source  
and thereby reduce or maybe even eliminate the costs of funding the deal. Thus, we have the following two definitions of the funding account: If rehypothecation of the posted collateral is allowed,
\begin{equation} \label{p4:eq:fund1}
F_t \triangleq {\bar V}_t - C_t - H_t , 
\end{equation}
and if such rehypothecation is forbidden, we have
\begin{equation} \label{p4:eq:fund2}
F_t \triangleq {\bar V}_t - H_t .
\end{equation}

By implication of \eqref{p4:eq:fund1} and \eqref{p4:eq:fund2} it is obvious that if the funding account $F_t>0$, the dealer needs to borrow cash to establish the hedging strategy at time $t$. Correspondingly, if the funding account $F_t<0$, the hedging strategy requires the dealer to invest surplus cash. Specifically, we assume the dealer enters a funding position on a discrete time-grid $\{t_1,\ldots,t_m\}$ during the life of the deal. Given two adjacent funding times $t_j$ and $t_{j+1}$, for $1\leq j \leq m-1$, the dealer enters a position in cash equal to $F_{t_j}$ at time $t_j$. At time $t_{j+1}$ the dealer redeems the position again and either returns the cash to the funder if it was a long cash position and pays funding costs on the borrowed cash, or he gets the cash back if it was a short cash position and receives funding benefits as interest on the invested cash. We assume that these funding costs and benefits are determined at the start date of each funding period and charged at the end of the period. 

Without any loss of generality, the contracts used by the investor to fund the deal can be introduced as adapted price processes. Let $P^{f^+}_t(T)$ represent the price of a borrowing contract measurable at $t$ where the dealer pays one unit of cash at maturity $T>t$, and let  $P^{f^-}_t(T)$ be the price of a lending contract where the dealer receives one unit of cash at maturity. Moreover, the corresponding accrual rates are given by
\[
f^\pm_t(T) \triangleq \frac{1}{T-t}\left(\frac{1}{P^{f^\pm}_t(T)}-1\right).
\]%
In other words, if the hedging strategy of the deal requires borrowing cash, this can be done at the funding frate $f^+$, while surplus cash can be invested at the lending rate $f^-$.
We define the effective funding rate ${\tilde f}_t$ faced by the dealer as
\begin{equation}
\label{p4:eq:ftilda}
{\tilde f}_t(T) \triangleq f^-_t(T) \ind{F_t<0} + f^+_t(T) \ind{F_t>0} \,.
\end{equation}
Following this, the sum of discounted cash-flows from funding the hedging strategy during the life of the deal is equal to
\begin{equation}
\label{p4:eq:fundinggeneral}
\varphi(t,T\wedge\tau;F) \triangleq \sum_{j=1}^{m-1} \ind{t\leqslant t_j< (T\wedge\tau)} D(t,t_j) F_{t_j} \left( 1 - \frac{P_{t_j}(t_{j+1})}{P^{\tilde f}_{t_j}(t_{j+1})} \right) \,,
\end{equation}%
where the zero-coupon bond corresponding to the effective funding rate is defined as $P^{\tilde f}_t(T) \triangleq [1+(T-t){\tilde f}_t(T) ]^{-1}$. This is, strictly speaking, a discounted payout and the funding cost or benefit at time $t$ is obtained by taking the risk neutral expectation of the above cash-flows. 

As before with collateral costs, we may rewrite the cash flows for funding as a first order approximation in continuously compounded rates $\tilde{f}$ and $r$ associated to the relevant bonds. We obtain
\begin{equation}
\label{p4:eq:fundinggeneral2}
\varphi(t,T\wedge\tau;F) \approx \sum_{j=1}^{m-1} \ind{t\leqslant t_j< (T\wedge\tau)} D(t,t_j) F_{t_j} \alpha_j  \left( r_{t_{j}}(t_{j+1}) - \tilde{f}_{t_{j}}(t_{j+1}) \right) \,,
\end{equation}%

We should also mention that, occasionally, we may include the effects of repo markets or stock lending in our framework. In general, we may borrow / lend the cash needed to establish $H$ from / to our treasury, and we may then use the risky asset in $H$ for repo or stock lending / borrowing in the market. This means that we could include the funding costs and benefits coming from this use of the risky asset. In this paper we assume that the bank Treasury automatically recognizes this benefit / cost to us at the same rate $\tilde{f}$ used for cash, but for a more general analysis involving repo rate $\tilde{h}$ see for example \cite{pallavicini2012funding}. 
%See Section \ref{p4:sec:extensions} below for a quick summary.

The particular positions entered by the dealer to either borrow or invest cash according to the sign and size of the funding account depend on the bank's liquidity policy. In the following we discuss two possible cases: One where the dealer can fund at rates set by the bank's treasury department, and another where the dealer goes to the market directly and funds his trades at the prevailing market rates. As a result, the funding rates and therefore the funding effect on the price of a derivative deal depends intimately on the chosen liquidity policy.
%Different levels of aggregration may be taken by the bank in setting its liquidity policy. 
%In the micro approach, we refer to this as case (a), the funding rates are determined at deal level. This means that the rate $f^+$ the desk can borrow funds at may be different from the rate $f^-$ at which it can invest funds. Moreover, these rates may differ across deals dedpending on the deals' notional, maturity structures, dealer-client relationship, and so forth. Increasing aggregration, in case (b), an average borrowing rate $f^+$ and an average $f^-$ lending rate are applied to all deals. Finally, in the large-pool approach refered to as case (c), the borrowing and lending rates are not only the same across all deals, they are also equal, i.e. $f^+=f^-$.

\paragraph{Treasury Funding}

If the dealer funds the hedge through the bank's treasury department, the treasury determines the funding rates $f^\pm$ faced by the dealer, often assuming average funding costs and benefits across all deals. This leads to two curves as functions of maturity; one for borrowing funds $f^+$ and one for lending funds $f^-$.
After entering a funding position $F_{t_j}$ at time $t_j$, the dealer faces the following discounted cash-flow 
\[
\Phi_j(t_j,t_{j+1};F) \triangleq  - N_{t_j} D(t_j,t_{j+1}) \,,
\]%
with
\[
N_{t_j} \triangleq \frac{F^-_{t_j}}{P^{f^-}_{t_j}(t_{j+1})} + \frac{F^+_{t_j}}{P^{f^+}_{t_j}(t_{j+1})} \,.
\]%
Under this liquidity policy, the treasury -- and not the dealer himself -- is in charge of debt valuation adjustments due to funding-related positions. Also, being entities of the same institution, both the dealer and the treasury disappear in case of default of the institution without any further cash-flows being exchanged and we can neglect the effects of funding in this case. So, when default risk is considered, this leads to following definition of the funding cash flows
\[
{\bar \Phi}_j(t_j,t_{j+1};F) \triangleq \ind{\tau>t_j} \Phi_j(t_j,t_{j+1};F) \,.
\]%
Thus, the risk-neutral price of the cash-flows due to the funding positions entered at time $t_j$ is
\[
\Ex{t_j}{{\bar \Phi}_j(t_j,t_{j+1};F)} = - \ind{\tau>t_j} \left( F^-_{t_j} \frac{P_{t_j}(t_{j+1})}{P^{f^-}_{t_j}(t_{j+1})} + F^+_{t_j} \frac{P_{t_j}(t_{j+1})}{P^{f^+}_{t_j}(t_{j+1})} \right) \,.
\]%
If we consider a sequence of such funding operations at each time $t_j$ during the life of the deal, we can define the sum of cash-flows coming from all the borrowing and lending positions opened by the dealer to hedge the trade up to the first-default event
\begin{align*}
\varphi(t,T\wedge\tau;F)
&\triangleq\, \sum_{j=1}^{m-1} \ind{t\leqslant t_j<(T\wedge\tau)} D(t,t_j) \left( F_{t_j} + \Ex{t_j}{{\bar \Phi}_j(t_j,t_{j+1};F)} \right) \\ \nonumber
& = \, \sum_{j=1}^{m-1} \ind{t\leqslant t_j<(T\wedge\tau)} D(t,t_j) \left( F_{t_j} - F^-_{t_j} \frac{P_{t_j}(t_{j+1})}{P^{f^-}_{t_j}(t_{j+1})} - F^+_{t_j} \frac{P_{t_j}(t_{j+1})}{P^{f^+}_{t_j}(t_{j+1})} \right) \,.
\end{align*}
In terms of the effective funding rate, this expression collapses to \eqref{p4:eq:fundinggeneral}.

\paragraph{Market Funding}

If the dealer funds the hedging strategy in the market -- and not through the bank's treasury -- the funding rates are determined by prevailing market conditions and are often deal specific. This means that the rate $f^+$ the dealer can borrow funds at may be different from the rate $f^-$ at which funds can be invested. Moreover, these rates may differ across deals depending on the deals' notional, maturity structures, dealer-client relationship, and so forth. Similar to the liquidity policy of treasury funding, we assume a deal's funding operations are closed down in the case of default. Furthermore, as the dealer now operates directly on the market, he needs to include a DVA due to his funding positions when he marks-to-market his trading books. For simplicity, we assume that the funder in the market is default-free so no funding CVA needs to be accounted for. The discounted cash-flow from the borrowing or lending position between two adjacent funding times $t_j$ and $t_{j+1}$ is given by
\begin{align*}
{\bar\Phi}_j(t_j,t_{j+1};F)
 \triangleq \, & \ind{\tau>t_j} \ind{\tau_I>t_{j+1}} \Phi_j(t_j,t_{j+1};F) \\
 &\,  -  \ind{\tau>t_j} \ind{\tau_I<t_{j+1}} (\lgd_I \varepsilon^-_{F,\tau_I} - \varepsilon_{F,\tau_I}) D(t_j,\tau_I) \,,
\end{align*}%
where $\varepsilon_{F,t}$ is the close-out amount calculated by the funder on the dealer's default
\[
\varepsilon_{F,\tau_I} \triangleq - N_{t_j} P_{\tau_I}(t_{j+1}) \,.
\]%
To price this funding cash-flow, we take the risk-neutral expectation
\[
\Ex{t_j}{{\bar \Phi}_j(t_j,t_{j+1};F)} = - \ind{\tau>t_j} \left( F^-_{t_j} \frac{P_{t_j}(t_{j+1})}{P^{f^-}_{t_j}(t_{j+1})} + F^+_{t_j} \frac{P_{t_j}(t_{j+1})}{{\bar P}^{f^+}_{t_j}(t_{j+1})} \right) .
\]%
Here, the zero-coupon funding bond ${\bar P}^{f^+}_t(T)$ for borrowing cash is adjusted for the dealer's credit risk
\[
{\bar P}^{f^+}_t(T) \triangleq  \frac{P^{f^+}_t(T)}{\ExT{t}{T}{ \lgd_I\ind{\tau_I>T} + \rec_I } }\,,
\]%
where the expectation on the right-hand side is taken under the $T$-forward measure. Naturally, since the seniority could be different, one might assume a different recovery rate on the funding position than on the derivatives deal itself (see \cite{Crepey2011}). Extensions to this case are straightforward.

Next, summing the discounted cash-flows from the sequence of funding operations through the life the deal, we get
\begin{align*}
\varphi(t,T\wedge\tau;F)
&\triangleq \, \sum_{j=1}^{m-1} \ind{t\leqslant t_j<T\wedge\tau} D(t,t_j) \left( F_{t_j} + \Ex{t_j}{{\bar \Phi}_j(t_j,t_{j+1};F)} \right) \\ \nonumber
& =\, \sum_{j=1}^{m-1} \ind{t\leqslant t_j<T\wedge\tau} D(t,t_j) \left( F_{t_j} - F^-_{t_j} \frac{P_{t_j}(t_{j+1})}{P^{f^-}_{t_j}(t_{j+1})} - F^+_{t_j} \frac{P_{t_j}(t_{j+1})}{{\bar P}^{f^+}_{t_j}(t_{j+1})} \right) \,.
\end{align*}%
To avoid cumbersome notation, we will not explicitly write ${\bar P}^{f^+}$ in the sequel, but just keep in mind that when the dealer funds directly in the market then ${P}^{f^+}$ needs to be adjusted for funding DVA. Thus, in terms of the effective funding rate, we obtain \eqref{p4:eq:fundinggeneral}.

\section{General Pricing Equation under Credit, Collateral and Funding}\label{p4:sec:pricingEq}

In the previous section we analyzed the discounted cash-flows of a derivative deal and we developed a framework for consistent valuation of such deals under collateralized counterparty credit and funding risk. The arbitrage-free pricing framework is captured in the following theorem.

\begin{theorem}[\textbf{General Pricing Equation}]
\label{p4:th:thCFBVA}
The consistent arbitrage-free price ${\bar V}_t(C,F)$ of collateralized OTC derivative deals with counterparty credit risk and funding costs takes the form
\begin{align}
\label{p4:eq:CFBVA}
{\bar V}_t(C,F) =  \mathbb E_{t} & \left[ \, \pi(t,T\wedge\tau) + \gamma(t,T\wedge\tau;C) + \varphi(t,T\wedge\tau;F) \right. \\ \nonumber
  & \left. + \ind{t<\tau<T} D(t,\tau) \theta_\tau(C,\varepsilon) \right],
\end{align}%
where 
\begin{enumerate}
	\item $\pi(t,T\wedge\tau)$ is the discounted cash-flows from the contract's payoff structure up to the first-default event.
	\item $\gamma(t,T\wedge\tau;C)$ is the discounted cash-flows from the collateral margining procedure up to the first-default event and is defined in \eqref{p4:eq:gamma}.
	\item $\varphi(t,T\wedge\tau;F)$ is the discounted cash-flows from funding the hedging strategy up to the first-default event and is defined in \eqref{p4:eq:fundinggeneral}.
	\item $\theta_\tau(C,\varepsilon)$ is the on-default cash-flow with close-out amount $\varepsilon$ and is defined in \eqref{p4:eq:theta}.
\end{enumerate}.
\end{theorem}
If funding and collateral margining costs are discarded, while collateral is retained for loss reduction at default, this pricing equation collapses to the formula derived in \cite{BrigoCapponiPallaviciniPapatheodorou} for the price of a derivative under bilateral counterparty credit risk. If further collateral guarantees are dropped, the formula reduces to the bilateral credit valuation formula in \cite{BrigoCapponi2010}.

While the pricing equation is conceptually clear -- we simply take the expectation of the sum of all discounted cash-flows of the deal under the risk-neutral measure -- solving the equation poses a recursive, non-linear problem. The future paths of the effective funding rate $\tilde f$ depend on the future signs of the funding account $F$, i.e. whether we need to borrow or lend cash on each future funding date. At the same time, through the relations \eqref{p4:eq:fund1} and \eqref{p4:eq:fund2}, the future sign and size of the funding account $F$ depend on the adjusted price $\bar V$ of the deal which is the quantity we are trying to compute in the first place. One crucial implication of this recursive structure of the pricing problem is the fact that FVA is generally not just an additive adjustment term, in contrast to CVA and DVA. More importantly, the conjecture identifying the DVA of a deal with its funding is not appropriate in general. Only in the unrealistic setting where the dealer can fund an uncollateralized trade at equal borrowing and lending rates, i.e. $f^+=f^-$, do we achieve the additive structure often assumed by practitioners. If the trade is collateralized, we need to impose even further restrictions as to how the collateral is linked to price of the trade $\bar V$.

\begin{remark}{\bf The law of one price.}
On the theoretical side, the pricing equation shakes the foundation of the celebrated Law of One Price prevailing in classical derivatives pricing. Clearly, if we assume no funding costs, the dealer and counterparty agree on the price of the deal as both parties can -- at least theoretically -- observe the credit risk of each other through CDS contracts traded in the market and the relevant market risks, thus agreeing on CVA and DVA. In contrast, introducing funding costs, they will not agree on the FVA for the deal due to asymmetric information. The parties cannot observe each others' liquidity policies nor their respective funding costs associated with a particular deal. As a result, the value of a deal position will not generally be the same to the counterparty as to the dealer just with opposite sign. In principle, this should mean that the dealer and the counterparty would never close the trade, but in practice trades are executed as a simple consequence of the fundamental forces of supply and demand. Nevertheless, among dealers it is the general belief that funding costs were one of the main factors driving the bid-ask spreads wider during the recent financial crisis. 
\end{remark}

Finally, as we adopt a risk-neutral valuation framework, we implicitly assume the existence of a risk-free interest rate. Indeed, since the valuation adjustments are included as additional cash-flows and not as ad-hoc spreads, all the cash-flows are discounted by the risk-free discount factor $D(t,T)$ in \eqref{p4:eq:CFBVA}. Nevertheless, the risk-free rate is merely an instrumental variable of the general pricing equation. We clearly distinguish market rates from the theoretical risk-free rate avoiding the false claim that the over-night rates (e.g., EONIA) are risk free. In fact, as we will show in continuous time, if the dealer funds the hedging strategy of the trade through cash accounts available to him -- whether as rehypothecated collateral or funds from the treasury, repo market, etc. -- the risk-free rate vanishes from the pricing equation.

\subsection{Discrete-time Solution}

Our purpose here is to turn the recursive pricing equation \eqref{p4:eq:CFBVA} into a set of iterative equations that can be solved by least-squares Monte Carlo methods. These methods are already standard in CVA and DVA calculations (\cite{BrigoPallavicini2007}). To this end, we introduce the auxiliary function 
\begin{equation}
\label{p4:eq:barpi}
{\bar\pi}(t_j,t_{j+1};C) \triangleq  \pi(t_j,t_{j+1}\wedge\tau) + \gamma(t_j,t_{j+1}\wedge\tau;C) + \ind{t_j<\tau<t_{j+1}} D(t_j,\tau) \theta_\tau(C,\varepsilon)
\end{equation}
which defines the cash-flows of the deal occurring between time $t_j$ and $t_{j+1}$ adjusted for collateral margining costs and default risks. We stress the fact that the close-out amount used for calculating the on-default cash-flow still refers to a deal with maturity $T$.
If we then solve pricing equation \eqref{p4:eq:CFBVA} at each funding date $t_j$ in the time-grid $\{t_1,\ldots,t_n=T\}$, we obtain the deal price $\bar V$ at time $t_j$ as a function of the deal price on the next consecutive funding date $t_{j+1}$
\begin{align*}
{\bar V}_{t_j} =  \Ex{t_j}{ {\bar V}_{t_{j+1}} D(t_j,t_{j+1}) + {\bar\pi}(t_j,t_{j+1};C) } + \ind{\tau>t_j}
                         \left( F_{t_j} - F^-_{t_j} \frac{P_{t_j}(t_{j+1})}{P^{f^-}_{t_j}(t_{j+1})}
                             - F^+_{t_j} \frac{P_{t_j}(t_{j+1})}{P^{f^+}_{t_j}(t_{j+1})}
                             \right) ,
\end{align*}
where, by definition, ${\bar V}_{t_n}\triangleq 0$ on the final date $t_n$. Recall the definitions of the funding account in \eqref{p4:eq:fund1} if no rehypothecation of collateral is allowed and in \eqref{p4:eq:fund2} if rehypothecation is permitted, we can then solve the above equation for the positive and negative parts of the funding account. The outcome of this exercise is a discrete-time iterative solution of the recursive pricing equation, provided in the following theorem.

\begin{theorem}[Discrete-time Solution of the General Pricing Equation]
\label{p4:th:iter}
We may solve the full recursive pricing equation in Theorem \ref{p4:th:thCFBVA} as a set of backward-iterative equations on the time-grid $\{t_1,\ldots,t_n=T\}$ with ${\bar V}_{t_n}\triangleq 0$. For $\tau<t_j$, we have
$${\bar V}_{t_j}=0,$$
while for $\tau>t_j$, we have
\begin{enumerate}
\item[(i)] if re-hypothecation is forbidden:
\[
\left( {\bar V}_{t_j} - H_{t_j} \right)^\pm = P^{\tilde f}_{t_j}(t_{j+1}) \left( \ExT{t_j}{t_{j+1}}{ {\bar V}_{t_{j+1}} + \frac{{\bar\pi}(t_j,t_{j+1};C)-H_{t_j}}{D(t_j,t_{j+1})} } \right)^\pm ,
\]
\item[(ii)] if re-hypothecation is allowed:
\[
\left( {\bar V}_{t_j}- C_{t_j} - H_{t_j}\right)^\pm =  P^{\tilde f}_{t_j}(t_{j+1}) \left( \ExT{t_j}{t_{j+1}}{ {\bar V}_{t_{j+1}} + \frac{{\bar\pi}(t_j,t_{j+1};C)-C_{t_j}-H_{t_j}}{D(t_j,t_{j+1})} } \right)^\pm ,
\]
\end{enumerate}
where the expectations are taken under the $\mathbb Q^{t_{j+1}}$-forward measure.
\end{theorem}
The $\pm$ sign in the theorem is supposed to stress the fact that the sign of the funding account, which determines the effective funding rate, depends on the sign of the conditional expectation. 
Further intuition may be gained by going to continuous time which is the case we will now turn to.

\subsection{Continuous-time Solution}\label{p4:sec:cont_time}

Let us consider a continuous-time approximation of the general pricing equation. This implies that collateral margining, funding, and hedging strategies are executed in continuous time. Moreover, we assume that rehypothecation is allowed, but similar results hold if this is not the case. By taking the time limit, we have the following expressions for the discounted cash-flow streams of the deal
\begin{equation*}
\pi(t,T\wedge\tau)=\int^{T\wedge\tau}_t \pi(s,s+ds) D(t,s) , \qquad \gamma(t,T\wedge\tau; C) =\int^{T\wedge\tau}_t (r_s-\tilde c_s) C_s D(t,s) ds ,
\end{equation*}
\begin{equation*}
\varphi(t,T\wedge\tau; F) =\int^{T\wedge\tau}_t (r_s-\tilde f_s) F_s D(t,s) ds , %\left[\bar V_s(C,F) - C_s -H_s\right]
\end{equation*}
where as before $\pi(t,t+dt)$ is the pay-off coupon process of the derivative contract and $r_t$ is the risk-free rate. These equations can also be immediately derived by looking at the approximations given in Equations (\ref{p4:eq:gamma2}) and (\ref{p4:eq:fundinggeneral2}).

Then, putting all the above terms together with the on-default cash flow as in Theorem \ref{p4:th:thCFBVA}, the recursive pricing equation yields
\begin{eqnarray}\label{eq:conttimeFVA}
{\bar V}_t  &=&\, \int_t^T \Ex{t}{ \left( \ind{s<\tau} \pi(s,s+ds) + \ind{\tau\in ds} \theta_s(C,\varepsilon) \right) D(t,s) }  \\ \nonumber
                &+&  \int_t^T  \Ex{t}{ \ind{s<\tau} ( r_s - {\tilde c}_s ) C_s D(t,s) }  ds  +  \int_t^T \Ex{t}{ \ind{s<\tau} ( r_s - {\tilde f}_s ) F_s   } D(t,s)  ds .
\end{eqnarray}
By recalling Equation (\ref{eq:cvadvacloseout}), we can write the following 
\begin{proposition}\label{prop:split}
The value $\bar{V}_t$ of the claim under credit gap risk, collateral and funding costs can be written as
\begin{equation}\label{eq:valuationsplit} \bar{V}_t = V_t - \mbox{CVA}_t + \mbox{DVA}_t + \mbox{LVA}_t + \mbox{FVA}_t
\end{equation}
where $V_t$ is the price of the deal when there is no credit risk, no collateral, and no funding costs; LVA is a liquidity valuation adjustment accounting for the costs/benefits of collateral margining; FVA is the funding cost/benefit of the deal hedging strategy, and CVA / DVA are the familiar credit and debit valuation adjustments after collateralization. These different adjustments can be written by rewriting Formula (\ref{eq:conttimeFVA}). One obtains
\begin{equation}\label{eq:vriskfree} V_t  =  \int_t^T \,\Exx_t \bigg{\{}  D(t,s) 1_{\{\tau > s\}} \bigg{[} \pi(s,s+ds) + 1_{\{\tau \in ds \}} \varepsilon_s \bigg{]} \bigg{\}}  
\end{equation}
and the valuation adjustments
\begin{align*}
\mbox{CVA}_t &= -\int_t^T \,\Exx \bigg{\{}  D(t,s) 1_{\{\tau > s\}} \big{[}  -  \ind{s=\tau_C<\tau_I}  \Pi_{\mbox{\footnotesize CVAcoll}}(s)  \big{]} \bigg{\}} du\\
\mbox{DVA}_t &= \int_t^T \,\Exx \bigg{\{}  D(t,s) 1_{\{\tau > s\}} \big{[}   \ind{s=\tau_I<\tau_C}  \Pi_{\mbox{\footnotesize DVAcoll}}(s) \big{]}  \bigg{\}} du \\
\mbox{LVA}_t &= \int_t^T \,\Exx_t \bigg{\{}  D(t,s) 1_{\{\tau > s\}} ( r_s - \tilde{c}_s ) C_s  \bigg{\}} ds \\
\mbox{FVA}_t &=  \int_t^T \,\Exx \bigg{\{}  D(t,s) 1_{\{\tau > s\}} \big{[}  ( r_s - \tilde{f}_s  ) F_s  
\big{]} \bigg{\}} ds 
\end{align*}
As usual, CVA and DVA are both positive, while LVA and FVA can be either positive or negative. Notice that if $\tilde{c}$ equals the risk free rate LVA vanishes. Similarly, FVA vanishes if the funding rate $\tilde{f}$ is equal to the risk-free rate.  
\end{proposition}

The proof is immediate. Notice that (\ref{eq:vriskfree}) simplifies further under a risk-free close-out
\[ \varepsilon_\tau = V_\tau . \]
Indeed, in such a case the presence of $\tau$ does not alter the present value. In fact, one can show that since the unwinding at $\tau$ happens at the fair price $V_\tau$ and with recovery one, this is equivalent, in terms of valuation at time $t$, to valuing the whole deal
\[ V_t =  \int_t^T \,\Exx_t \bigg{\{}  D(t,s) \pi(s,s+ds) \bigg{\}} . \]

\begin{remark}{\bf Separability?} 
As pointed out earlier and in \cite{pallavicini2011funding}, valuation Formula (\ref{eq:valuationsplit}) is not really splitting risk components in different terms. For example, $\tilde{f}$ needs to know future signs of $F$ and hence future $\bar{V}$'s. Such future $\bar{V}$'s depend on all risks together, and so does $\tilde{f}$. Hence interpreting the $\tilde{f}$-dependent term FVA as a pure funding adjustment is misleading. Similarly, if we adopt a replacement close-out at default where $\varepsilon_\tau = \bar{V}_{\tau-}$, then the CVA and DVA terms will depend on future $\bar{V}$, and hence on all other risks as well. Therefore, CVA is no longer a pure credit valuation adjustment. If enforcing the above separation a posteriori  after solving the total equation for $\bar{V}$, then one has to be careful in interpreting this as a real split of risks. A further point concerns the presence of the short rate $r_t$ in the terms above.     
Since $r$ is a theoretical rate with no direct market counterpart, this decomposition is not ideal. To implement the pseudo-decomposition, one would have to proxy $r$ with a real market rate, such as for example an overnight rate. 
\end{remark}

By recalling that $\bar{V}_t = F_t + H_t + C_t$, we may rewrite Equation (\ref{eq:conttimeFVA}) as

\begin{align*}
{\bar V}_t  = &\, \int_t^T \Ex{t}{ \left( \ind{s<\tau} \pi(s,s+ds) + \ind{\tau\in ds} \theta_s(C,\varepsilon) \right) D(t,s) }  \\
                &\, +  \int_t^T  \Ex{t}{ \ind{s<\tau} ( {\tilde f}_s - {\tilde c}_s ) C_s D(t,s) }  ds \\
                &\, +  \int_t^T \Ex{t}{ \ind{s<\tau} ( r_s - {\tilde f}_s ) \left( {\bar V}_s - H_s \right)D(t,s) } ds .
\end{align*}
For simplicity, we now adopt an immersion hypothesis and switch to the default-free market filtration $({\mathcal F}_t)_{t \ge 0}$. This step implicitly assumes a separable structure of our complete filtration $({\mathcal G}_t)_{t \ge 0}$. In other words, ${\mathcal G}_t$ is generated by the pure default-free market filtration ${\mathcal F}_t$ and by the filtration generated by all the relevant default times monitored up to $t$. We are also assuming that the basic portfolio cash flows $\pi(0,t)$ are ${\cal F}_t$ measurable and that default times of all parties are conditionally independent given filtration ${\cal F}$ (see for example \cite{BieleckiRutkowski2002}, or \cite{BrigoPallaviciniJFE} for the full details in the present setup). We assume all needed technical conditions to be satisfied. This allows us to rewrite the previous price equation as
\begin{align*}
{\bar V}_t
 = &\, \ind{\tau>t} \int_t^T \, \Ex{t}{ \left( \pi(s,s+ds) + \lambda_s \theta_s(C,\varepsilon) ds \right) D(t,s;r+\lambda) \big| {\mathcal F}} \\
&\, +  \ind{\tau>t} \int_t^T  \, \Ex{t}{ ( {\tilde f}_s - {\tilde c}_s ) C_s D(t,s;r+\lambda) \big| {\mathcal F} } ds \\
&\, +  \ind{\tau>t} \int_t^T \, \Ex{t}{ ( r_s - {\tilde f}_s ) \left( {\bar V}_s-H_s \right) D(t,s;r+\lambda) \big| {\mathcal F} } ds
\end{align*}%
where $\lambda_t$ is the first-to-default intensity and the discount factor is defined as $D(t,s;\xi) \triangleq e^{-\int_t^s \xi_u du }$. We further denote $\Ex{t}{\cdot |{\cal F}}$ the conditional expectation with respect to ${\cal F}_t$. Assuming the relevant technical conditions are satisfied, the Feynman-Kac theorem now allows us to formally write down the corresponding pre-default partial differential equation (PDE) of the pricing problem. The formal proof is the same as that of Feynman-Kac for the linear case and relies on the stochastic integral term being a martingale and applying It\^o's formula twice. Further, we need to assume that the underlying market risk factors are Markov processes with infinitesimal generator $\mathcal L_t$, that the risk-free rate $r_t$ is bounded, and that we have sufficient smoothness of the price function $\bar V$. Hence, for $\tau> t$, we get the PDE
\begin{equation}\label{p4:eq:PDE}
( \partial_t - {\tilde f}_t - \lambda_t + {\mathcal L}_t ) {\bar V}_t - ( r_t - {\tilde f}_t ) H_t + ( {\tilde f}_t - {\tilde c}_t ) C_t + \lambda_t \theta_t + \pi_t = 0
\end{equation}
with boundary condition
\begin{equation}\label{p4:eq:PDEbc}
{\bar V}_T = 0
\end{equation}
and where 
\begin{equation}
 \pi_t  = \lim_{h\downarrow 0} \frac{\pi(t,t+h)}{h},
\end{equation}
where the single argument in $\pi$ avoids confusion on which $\pi$ we are referring to, or, more informally, 
\[ \pi_t dt = \pi(t,t+dt) \]
for small $dt$, where $\pi_t$ is assumed ${\cal F}_t$ adapted. 
In \cite{BrigoPallaviciniJFE} a BSDE derivation of the PDE is provided, avoiding the Feynman Kac argument. 
 
If we, furthermore, assume that the underlying risk factors are diffusions and consider delta-hedging of the deal value $\bar V$ in continuous time, the generator $\mathcal L$ can be expanded in terms of first and second order operators
\[
{\mathcal L}_t {\bar V}_t \triangleq \left( {\mathcal L}_t^1 + {\mathcal L}_t^2 \right) {\bar V}_t \triangleq r_t H_t + {\mathcal L}_t^2 {\bar V}_t.
\]
As a result, for $\tau> t$, the pricing PDE reduces to
\begin{equation}\label{eq:predefaultpdedelta}
( \partial_t - {\tilde f}_t - \lambda_t + {\mathcal L}_t^{\tilde f} ) {\bar V}_t + ( {\tilde f}_t - {\tilde c}_t ) C_t + \lambda_t \theta_t +  \pi_t = 0,
\end{equation}
where ${\mathcal L}_t^{\tilde f} {\bar V}_t \triangleq {\tilde f}_t H_t + {\mathcal L}_t^2 {\bar V}_t$.
Observe that the pre-default PDE no longer depends on the risk-free rate $r_t$. This equation may be solved numerically as in \cite{Crepey2012a}. On the other hand, if we apply the Feynman-Kac theorem again -- this time going from the pre-default PDE to the pricing expectation -- and integrate by parts, we arrive at the following result
\begin{theorem}[\textbf{Continuous-time Solution of the General Pricing Equation}]
\label{p4:th:fund_hedge}
If we assume collateral rehypothecation and delta-hedging, we can solve the iterative equations of Theorem~\ref{p4:th:iter} in continuous time. We obtain
\begin{equation}\label{eq:fundingafterFK}
  \bar{V}_t
 =  \int_t^T \,\Exx^{\tilde{f}}\{  D(t,u;\tilde{f}+\lambda) [ \pi_u + \lambda_u \theta_u + ( \tilde{f}_u - \tilde{c}_u ) C_u ]|{\cal F}_t \} du  
\end{equation}
where expectations are taken under the pricing measure $\mathbb{Q}^{\tilde f}$ for which the underlying risk factors grow at the rate ${\tilde f}$ if no dividend is paid.
\end{theorem}
Theorem \ref{p4:th:fund_hedge} decomposes the deal price $\bar V$ into three intuitive terms. The first term is the value of the deal cash flows, discounted at funding plus credit. The second term is the price of the on-default cash-flow in excess of the collateral, which includes the CVA and DVA of the deal after collateralization. The last term collects the cost of collateralization. In addition, we see that any dependence on the hedging strategy $H$ can be dropped by taking all expectations under the pricing measure $\mathbb Q^{\tilde f}$.
At this point it is very important to appreciate once again that $\tilde{f}$ depends on $F$, and hence on~$V$.  

\begin{remark}{\bf (Deal dependent pricing measure, local risk neutral measures).} Since the pricing measure depends on $\tilde{f}$ which in turn depends on the very value $\bar{V}$ we are trying to compute, we have that the pricing measure becomes deal dependent. Every deal or portfolio has a different pricing measure. 
\end{remark} 

Finally, we stress once again a very important invariance result that first appeared in \cite{pallavicini2012funding} and \cite{BrigoMoriniPallavicini2012}.

\begin{theorem}{\bf (Invariance of the valuation equation with respect to the short rate~$r_t$)}. Equations 
(\ref{eq:predefaultpdedelta}) or (\ref{eq:fundingafterFK}) for valuation under credit, collateral and funding costs are completely governed by market rates; there is no dependence on a risk-free rate $r_t$. Whichever initial process is postulated for $r$, the final price is invariant to it.  This confirms our earlier conjecture that the risk-free rate is merely an instrumental variable of our valuation framework and we do not, in fact, need to know the value of such a rate.
\end{theorem}
The proof is immediate by inspection.

\section{Numerical Results: Extending the Black-Scholes Analysis}\label{p4:sec:numerics}

This section provides a numerical study of the valuation framework outlined in the previous sections. We investigate the impact of funding risk on the price of a derivatives deal under default risk and collateralization. To this end, we propose a least-squares Monte Carlo algorithm inspired by the simulation methods of \cite{carriere1996valuation}, \cite{longstaff2001valuing}, \cite{tilley1993valuing}, and \cite{tsitsiklis2001regression} for pricing American-style options. As the purpose is to understand the fundamental implications of funding risk, we focus on relatively simple deal positions in European call options. However, the Monte Carlo method we propose below can be applied to more complex derivative contracts, including derivatives with bilateral payments.

\subsection{Monte Carlo Algorithm}

Recall the recursive structure of the general pricing equation: The deal price depends on the funding decisions, while the funding strategy depends on the future price itself. 
The intimate relationship among the key quantities makes the pricing problem computationally challenging.

We consider $K$ default scenarios during the life of the deal -- either obtained by simulation, bootstrapped from empirical data, or assumed in advance. For each first-to-default time $\tau$ corresponding to a default scenario, we compute the price of the deal $\bar V$ under collateralization, close-out netting and funding costs.
The first step of our simulation method entails simulating a large number of sample paths $N$ of the underlying risk factors $X$. We simulate these paths on the time-grid $\{t_1,\ldots,t_m =T^*\}$ with step size $\Delta t=t_{j+1} -t_j$ from the assumed dynamics of the risk factors. $T^*$ is equal to the final maturity $T$ of the deal or the consecutive time-grid point following the first-default time $\tau$, whichever occurs first. 
For simplicity, we assume the time periods for funding decisions and collateral margin payments coincide with the simulation time grid. 

Given the set of simulated paths, we solve the funding strategy recursively in a dynamic programming fashion. 
Starting one period before $T^*$, we compute for each simulated path the funding decision $F$ and the deal price $\bar V$ according to the set of backward-inductive equations of Theorem \ref{p4:th:iter}. The algorithm then proceeds recursively until time zero. Ultimately, the total price of the deal is computed as the probability weighted average of the individual prices obtained in each of the $K$ default scenarios.

The conditional expectations in the backward-inductive funding equations are approximated by across-path regressions based on least squares estimation similar to \cite{longstaff2001valuing}. We regress the present value of the deal price at time $t_{j+1}$, the adjusted payout cash flow between $t_{j}$ and $t_{j+1}$, the collateral account and funding account at time $t_{j}$ on basis functions $\psi$ of realizations of the underlying risk factors at time $t_{j}$ across the simulated paths. To keep notation simple, let us assume that we are exposed to only one underlying risk factor, e.g. a stock price. Extensions to higher dimensions are straightforward. Specifically, the conditional expectations in the iterative equations of Theorem \ref{p4:th:iter}, taken under the risk-neutral measure, are equal to
\begin{equation}
\mathbb{E}_{t_j}\left[\Xi_{t_j}(\bar{V}_{t_{j+1}})\right] = \theta_{t_{j}}' \,\psi(X_{t_{j}}),
\end{equation}
where we have defined $\Xi_{t_j}(\bar{V}_{t_{j+1}})\triangleq D(t_j,t_{j+1})\bar{V}_{t_{j+1}} + \bar{\pi}(t_j,t_{j+1};C) - C_{t_j} -H_{t_j}$. Note the $C_{t_j}$ term drops out if rehypothecation is not allowed.
The usual least-squares estimator of $\theta$ is then given by
\begin{equation}
\hat\theta_{t_{j}} \triangleq \left[\psi(X_{t_{j}})\psi(X_{t_{j}})'\right]^{-1}\psi(X_{t_{j}})\, \Xi_{t_j}(\bar{V}_{t_{j+1}}).
\end{equation}
Orthogonal polynomials such as Chebyshev, Hermite, Laguerre, and Legendre may all be used as basis functions for evaluating the conditional expectations. We find, however, that simple power series are quite effective and that the order of the polynomials can be kept relatively small. In fact, linear or quadratic polynomials, i.e. $\psi(X_{t_{j}})= (\textbf{1}, X_{t_{j}}, X_{t_{j}}^2)'$, are often enough. 

Further complexities are added, as the dealer may -- realistically -- decide to hedge the full deal price $\bar V$. Now, the hedge $H$ itself depends on the funding strategy through $\bar V$, while the funding decision depends on the hedging strategy. This added recursion requires that we solve the funding and hedging strategies simultaneously. 
For example, if the dealer applies a delta-hedging strategy
\begin{align}\label{p4:eq:deltahedging}
H_{t_j}=\frac{\partial \bar V}{\partial X} \Big\lvert_{t_j} \, X_{t_j} \approx \frac{\bar V_{t_{j+1}} - (1+\Delta t_j\tilde f_{t_{j}}) \bar V_{t_{j}}}{X_{t_{j+1}}  -(1+\Delta t_j\tilde f_{t_{j}}) X_{t_{j}}}\, X_{t_j},
\end{align}
we obtain, in the case of rehypothecation, the following system of non-linear equations 
\begin{equation}
\begin{cases} F_{t_j}  - \frac{P^{\tilde f}_{t_j}(t_{j+1})}{P_{t_j}(t_{j+1})} \,\mathbb{E}_{t_j}\left[\Xi_{t_j}(\bar{V}_{t_{j+1}})\right] = 0,  \\ 
H_{t_j} -\frac{\bar V_{t_{j+1}} - (1+\Delta t_j\tilde f_{t_{j}}) \bar V_{t_{j}}}{X_{t_{j+1}}  - (1+\Delta t_j\tilde f_{t_{j}}) X_{t_{j}}}\, X_{t_j} = 0,\\
\bar V_{t_j} = F_{t_j} + C_{t_j} + H_{t_j},\end{cases}
\end{equation}
where all matrix operations are on an element-by-element basis. 
An analogous result holds when rehypothecation of the posted collateral is forbidden.

Each period and for each simulated path, we find the funding and hedging decisions by solving this system of equations, given the funding and hedging strategies for all future periods until the end of the deal. We apply a simple Newton-Raphson method to solve the system of non-linear equations numerically, but instead of using the exact Jacobian, we approximate it by finite differences. As initial guess, we use the Black-Scholes delta position
$$H^0_{t_j} = \Delta^{BS}_{t_j}\, X_{t_j}.$$
The convergence is quite fast and only a small number of iterations are needed in practice. 
Finally, if the dealer decides to hedge only the risk-free price of the deal, i.e. the classic derivative price $V$, the pricing problem collapses to a much simpler one. The hedge $H$ no longer depends on the funding decision and can be computed separately and the numerical solution of the non-linear equation system can be avoided altogether.

%Finally, we note that the collateral margin procedure may well depend on $V$ but is assumed independent of the total deal price $\bar V$ including funding costs. T However, similar to the case when the hedge depends on the deal price, the algorithm can be extended along the same lines to $C$ being some function $\bar V$. 

Now we move on to analyzing numerical results on the impact of funding for both a long and a short position in a European call option. Before doing so, we fully specify our modeling setup.

\subsection{Market, Credit, and Funding Risk Specification}

We assume that the underlying stock $S_t$ evolves according to a geometric Brownian motion $dS_t = rS_t dt + \sigma S_t dW_t$ where $W$ is a standard Brownian motion under the risk neutral measure. The risk-free interest rate $r$ is $100$ bps, the volatility $\sigma$ is $25\%$, and the current price of the underlying is $S_0=100$. 
The European call option is in-the-money and has strike $K=80$. The maturity $T$ of the deal is 3 years and, in the full case,  we assume that the investor delta-hedges the deal according to \eqref{p4:eq:deltahedging}. The usual default-free funding-free and collateral-free Black-Scholes price $V_0$  of the call option deal is given by
\[  V_t = S_t \Phi(d_1(t)) - K e^{-r (T-t)}  \Phi(d_2(t)), \ \ d_{1,2} = \frac{\ln(S_t/K) +  (r \pm \sigma^2 / 2)(T-t)  }{\sigma \sqrt{T-t}}    \]
for $t=0$ and is 
\[ V_0 = 28.9\]  
with our choice of inputs. As usual, $\Phi$ is the cumulative distribution function of the standard normal random variable. In the usual setting the hedge would not be $\eqref{p4:eq:deltahedging}$ but a classical delta-hedging strategy based on $\Phi(d_1(t))$. 

We consider two simple discrete probability distributions of default. Both parties of the deal are considered default risky but can only default at year 1 or at year 2. The localized joint default probabilities are provided in the matrices below. The rows denote the default time of the investor, while the columns denote the default times of the counterparty. For example, in matrix $D_{\text{low}}$ the event $(\tau_I=2yr,\tau_C=1yr)$ has a 3\% probability and the first-to-default time is 1 year. Simultaneous defaults are introduced and we determine the close-out amount by a random draw from a uniform distribution. If the random number is above 0.5, we compute the close-out as if the counterparty defaulted first, and vice versa.

For the first default distribution, we have a low dependence between the default risk of the counterparty and the default risk of the investor
\begin{equation}
D_{\text{low}} = \bordermatrix{~ & 1yr  & 2yr & n.d. \cr
                   1yr & 0.01 & 0.01 & 0.03 \cr
                  2yr & 0.03 & 0.01 & 0.05 \cr
                   n.d. & 0.07 & 0.09 & 0.70 \cr}\,,\qquad\quad \tau_K(D_{\text{low}})=0.21  %_{\text{Kendall}}
\end{equation}
where $n.d.$ means no default and $\tau_K$ denotes the rank correlation as measured by Kendall's tau. In the second case, we have a high dependence between the two parties' default risk
\begin{equation}
D_{\text{high}} = \bordermatrix{~ & 1yr  & 2yr & n.d. \cr
                   1yr & 0.09 & 0.01 & 0.01 \cr
                  2yr & 0.03 & 0.11 & 0.01 \cr
                   n.d. & 0.01 & 0.03 & 0.70 \cr}\,,\qquad\quad \tau_K(D_{\text{high}})=0.83
\end{equation}
Note also that the distributions are skewed in the sense that the counterparty has a higher default probability than the investor.
The loss given default is $50\%$ for both the investor and the counterparty and the loss on any posted collateral is considered the same. The collateral rates are chosen to be equal to the risk free rate.
We assume that the collateral account is equal to the risk-free price of the deal at each margin date, i.e. $C_{t} = V_{t}$. This is reasonable as the dealer and client will be able to agree on this price, in contrast to $\bar V_{t}$ due to asymmetric information. Also, choosing the collateral this way has the added advantage that the collateral account $C$ works as a control variate, reducing the variance of the least-squares Monte Carlo estimator of the deal price.

\subsection{Preliminary Analysis without Credit Risk and with Symmetric Funding Rates}

To provide some ball-park figures on the effect of funding risk, we first look at the case without default risk and without collateralization of the deal. We compare our Monte Carlo approach to three alternative (simplified) approaches:
\begin{itemize}
\item[(i)] Simple discounting of the risk-free Black-Scholes price with a symmetric funding rate $\hat{f} = f^+ = f^-$. We obtain
\[  V^{(i)}_t = e^{- \hat{f}  T} \left(S_t \Phi(d_1(t)) - K e^{-r (T-t)}  \Phi(d_2(t))\right),  \]
assuming a continuously compounded funding rate.
\item[(ii)] The Black-Scholes price where both discounting and the growth of the underlying happens at the symmetric funding rate
\[  V^{(ii)}_t = \left(S_t \Phi(g_1(t)) - K e^{-\hat{f} (T-t)}  \Phi(g_2(t))\right), \ \  
g_{1,2} = \frac{\ln(S_t/K) +  (\hat{f} \pm \sigma^2 / 2)(T-t)  }{\sigma \sqrt{T-t}}  \]
%\item[(iii)] Simple discounting of the forward price with the symmetric funding rate. This approach can be justified by the fact that the price of a deep out-of-the-money call will approximately equal that of a forward contract.
%\[  V^{(iii)}_t = S_t - K e^{-\hat{f} (T-t)}   \]
\item[(iii)] We use the above FVA formula in Proposition \ref{prop:split} with some approximations. Since in a standard Black scholes setting 
$F_t = - K e^{-r (T-t)}\Phi(d_2(t))$, we compute
\[ \mbox{FVA}^{(iv)} = ( r - \hat{f}  ) \int_0^T \,\Exx_0 \left\{  e^{-r s} [   F_s ]  \right\} ds 
= ( \hat{f} - r  )  K e^{-r T} \int_0^T \,\Exx_0 \left\{ \Phi(d_2(s))  \right\} ds  \]
\end{itemize}
We illustrate the three approaches in the case of an equity call option (long position). Moreover, let the funding valuation adjustment in each case be defined by $\mbox{FVA}^{(i,ii,iii)} = V^{(i,ii,iii)} - V$. Figure \ref{fig4} plots the resulting funding valuation adjustment with credit and collateral switched off under the three different approaches and under the full valuation approach. Recall that if the funding rate is equal to the risk-free rate, the value of the call option collapses to the Black-Scholes price and the funding valuation adjustment is zero.

\begin{figure}
\centering
\includegraphics[scale=0.60]{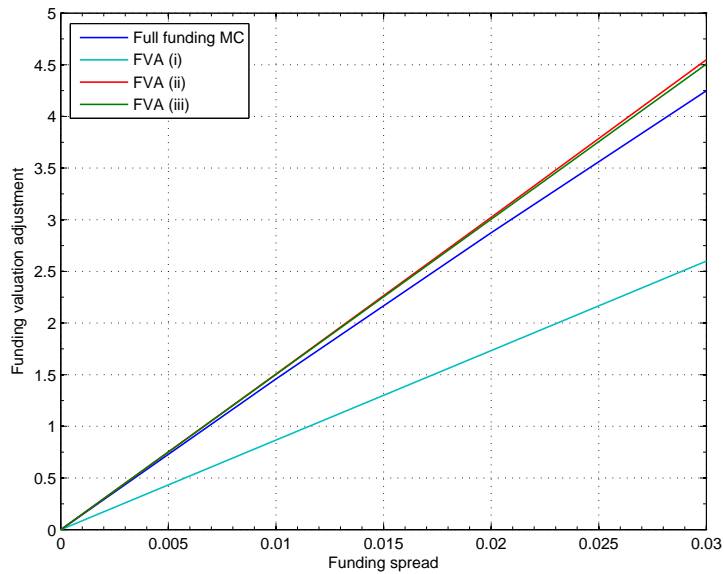}%[scale=0.5]
\caption{Funding valuation adjustment of a long call position as a function of symmetric funding spreads $s_f := \hat{f} - r$ with $\hat{f}:=f^+=f^-$. The adjustments are computed under the assumption of no default risk nor collateralization.}
\label{fig1}
\end{figure}

\begin{remark}{\bf (Current market practice for FVA).}
Looking at Figure \ref{fig1}, it is important to realize that at the time of writing this paper, most market players would adopt a methodology like (ii) or (iii) for a simple call option.  Even if borrowing or lending rates were different, most market players would average them and apply a common rate to borrowing and lending, in order to avoid non-linearities. We will discuss the approximation error entailed in this symmetrization later, when introducing the NVA. For the time being, we notice that method (iii) produces the same results as the quicker method (ii), that simply replaces the risk free rate by the funding rate. In the simple case of a long position in a call option without credit and collateral, and with symmetric borrowing and lending rates, we can show that this method is sound since it stems directly from our rigorous Formula (\ref{eq:fundingafterFK}). We also see that both methods (ii) and (iii) are quite close to the full numerical method we adopt. Occasionally, the industry may adopt methods such as (i), but this is not recommended, as we can see from the results. 
Overall industry-like methods such as (ii) or (iii) work well here, and there would be no need to implement the full machinery. However, once collateral and credit risk are in the picture, and once non-linearities due to replacement close-out at default and asymmetry in borrowing and lending are present, there is no way we can keep using something like (ii) or (iii) and we need to implement the full methodology. 
\end{remark}

\subsection{Full Analysis with Credit Risk, Collateral and Funding Costs}

Let us now switch on credit risk and consider collateralized deals. The recursive structure of our simulation method makes the pricing problem particularly demanding in terms of computational time, so we are forced to choose a relatively small number of sample paths. We use 1,000 sample paths but, fortunately, the presence of collateral as a control variate mitigates too large errors. In Tables \ref{p4:tab:numericsColl}-\ref{p4:tab:numericsCollRehyp} we conduct a ceteris paribus analysis of funding risk under counterparty credit risk and collateralization. Specifically, we investigate how the value of a deal changes for different values of the borrowing (lending) rate $f^+$ ($f^-$) while keeping the lending (borrowing) rate fixed to 100 bps. When both funding rates are equal to 100 bps the deal is funded at the risk-free rate and we are in the classical derivatives pricing setting.

\begin{remark}{\bf (Potential arbitrage).}
Note that if $f^+ < f^-$ arbitrage opportunities might be present, unless certain constraints are imposed on the funding policy of the treasury. Such constraints may look unrealistic and may be debated themselves from the point of view of arbitrageability, but since our point here is strictly to explore the impact of asymmetries in the funding equations we will still apply our framework to a few examples where $f^+ < f^-$.
\end{remark}

Table \ref{p4:tab:numericsColl} reports the impact of changing funding rates for a call position when the posted collateral may not be used for funding the deal, i.e. rehypothecation is not allowed. First, for the long position, increasing the lending rate $f^-$ while keeping the borrowing rate $f^+$ fixed causes an increase in the deal value. On the other hand, an increase in the borrowing rate while fixing the lending rate, decreases the value of the short position, i.e. the negative exposure of the investor increases. As a call option is just a one-sided contract, increasing the borrowing rate for a long position only has a minor impact. Recall that $F$ is defined as the cash account needed as part of the derivative replication strategy or, analogously, the cash account required to fund the hedged derivative position. To hedge a long call, the investor goes short in a delta position of the underlying asset and invests excess cash in the treasury at $f^-$. Correspondingly, to hedge the short position, the investor enters a long delta position in the stock and finances it by borrowing cash from the treasury at $f^+$, so changing the lending rate only has a small effect on the deal value. Finally, due to the presence of collateral, we observe an almost similar price impact of funding under the two different default distributions $D_{\text{low}}$ and $D_{\text{high}}$. 

\begin{table}
\centering\small
\begin{minipage}[t]{335px}\caption{Price impact of funding with default risk and collateralization}\label{p4:tab:numericsColl}
\end{minipage}
\begin{tabular}{lcccc}%p{80px} >{\raggedleft}p{80px} >{\raggedleft}p{80px} >{\raggedleft}p{80px} >{\raggedleft}p{80px}}
\toprule
\midrule
   & \multicolumn{2}{c}{Default risk, low$^{\text b}$}  & \multicolumn{2}{c}{Default risk, high$^{\text c}$}  \ \tabularnewline
 \cmidrule(r){2-3}\cmidrule(r){4-5}
 Funding$^{\text a}$  & \multicolumn{1}{c}{Long} 	 & \multicolumn{1}{c}{Short} & \multicolumn{1}{c}{Long} 	 & \multicolumn{1}{c}{Short} \tabularnewline
\midrule
\textit{Borrowing rate} $f^+$		&	& &	& \\
$\;\;\;$ $0$ bps & 29.36 (0.12) & -26.20 (0.17) & 29.67 (0.22) & -26.60 (0.36)	\\
$\;\;\;$ $100$ bps & 28.70 (0.15) & -28.72 (0.15) & 29.06 (0.21)  & -29.07 (0.21) \\
$\;\;\;$ $200$ bps & 28.05 (0.21)  & -31.37 (0.32) & 28.45 (0.22)  & 	-31.66 (0.25)		\\
$\;\;\;$ $300$ bps & 27.38 (0.29) & -34.26 (0.55) & 27.83 (0.23)  & 	-34.48 (0.46)	\\
$\;\;\;$ $400$ bps & 26.67 (0.38)  & -37.24 (0.86)  & 27.17 (0.26)  & -37.38 (0.80)		\\
\midrule
\textit{Lending rate} $f^-$		&	& &	& \\
$\;\;\;$ $0$ bps & 26.17 (0.18) & 	-29.38 (0.11) & 26.59 (0.36)  & 	-29.68 (0.22)	\\
$\;\;\;$ $100$ bps & 28.70 (0.15)  & 	-28.72 (0.15) & 29.06 (0.21)  & -29.07 (0.21)		\\
$\;\;\;$ $200$ bps & 31.37 (0.32)  & 	-28.07 (0.22) & 31.67 (0.25)  & 	-28.46 (0.22)		\\
$\;\;\;$ $300$ bps & 34.28 (0.55)  & 	-27.41 (0.30) & 34.51 (0.47)  & 	-27.85 (0.23)		\\
$\;\;\;$ $400$ bps & 37.28 (0.88)  & 	-26.69 (0.39) & 37.45 (0.82)  & 	-27.17 (0.26)	\\
\midrule
\bottomrule
\end{tabular}
\hspace{0.1cm}
\\ \begin{minipage}[t]{335px}\footnotesize{Standard errors of the price estimates are given in parentheses.\\
$^{\text a}$ Ceteris paribus changes in one funding rate while keeping the other fixed to 100 bps.\\ 
$^{\text b}$ Based on the joint default distribution $D_{\text{low}}$ with low dependence.\\
$^{\text c}$ Based on the joint default distribution $D_{\text{high}}$ with high dependence.\\ }
\end{minipage}
\end{table}

\begin{figure}
\centering
\includegraphics[scale=0.60]{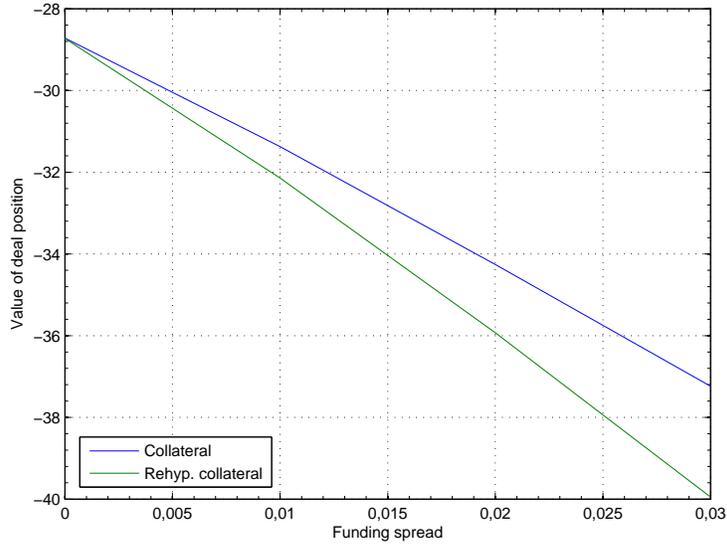}%[scale=0.5]
\caption{The value of a short call position for asymmetric funding spreads $s^+_f = f^+ - r$, i.e. fixing $f^-=r=0.01$ and varying $f^+\in(0.01, 0.02, 0.03, 0.04)$.}
\label{fig3}
\end{figure}

\begin{figure}
\centering
\includegraphics[scale=0.60]{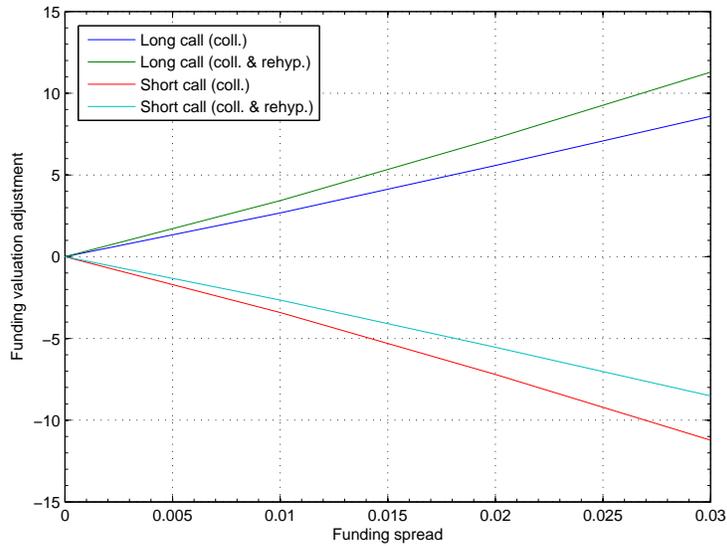}%[scale=0.5]
\caption{Funding valuation adjustment as a function of asymmetric funding spreads. The adjustments are computed under the presence of default risk and collateralization.}
\label{fig4}
\end{figure}

Finally, assuming cash collateral, we consider the case of rehypothecation and allow the investor and counterparty to use any posted collateral as a funding source. If the collateral is posted to the investor, this means it effectively reduces his costs of funding the delta-hedging strategy. As the payoff of the call is one-sided, the investor only receives collateral when he holds a long position in the call option. But as he hedges this position by short-selling the underlying stock and lending the excess cash proceeds, the collateral adds to his cash lending position and increases the funding benefit of the deal. Analogously, if the investor has a short position, he posts collateral to the counterparty and a higher borrowing rate would increase his costs of funding the collateral he has to post as well as his delta-hedge position. 
Table \ref{p4:tab:numericsCollRehyp} reports the results for the short and long positions in the call option when rehypothecation is allowed. 
Figure \ref{fig3} plots the values of collateralized short positions in the call option as a function of asymmetric funding spreads. In addition, Figure \ref{fig4} reports the corresponding FVA defined as the difference between the full funding-inclusive deal price and the full deal price but symmetric funding rates equal to the risk-free rate. Recall that the collateral rates are equal to the risk free rate, so the LVA collapses to zero in these examples.

%\begin{figure}
%\centering
%\includegraphics[scale=0.60]{fig2.eps}%[scale=0.5]
%\caption{The value of a long call position for asymmetric funding spreads $s^-_f = f^- - r$, i.e. fixing $f^+=r=0.01$ and varying $f^-\in(0.01, 0.02, 0.03, 0.04)$.}
%\label{fig2}
%\end{figure}

This shows that funding asymmetry matters even under full collateralization when there is no repo market for the underlying stock. In practice, however, the dealer cannot hedge a long call by shorting a stock he does not own. Instead, he would first borrow the stock in a repo transaction and then sell it in the spot market. Similarly, to enter the long delta position needed to hedge a short call, the dealer could finance the purchase by lending the stock in a reverse repo transaction. Effectively, the delta position in the underlying stock would be funded at the prevailing repo rate. Thus, once the delta hedge has to be executed through the repo market, there is no funding valuation adjustment (meaning any dependence on the funding rate $\tilde f$ drops out) given the deal is fully collateralized, but the underlying asset still grows at the repo rate. If there is no credit risk, this would leave us with the result of \cite{Piterbarg2010}. However, if the deal is not fully collateralized or the collateral cannot be rehypothecated, funding costs enter the picture even when there is a repo market for the underlying stock.

\begin{table}
\centering\small
\begin{minipage}[t]{335px}\caption{Price impact of funding with default risk, collateralization, and rehypothecation}\label{p4:tab:numericsCollRehyp}
\end{minipage}
\begin{tabular}{lcccc}%p{80px} >{\raggedleft}p{80px} >{\raggedleft}p{80px} >{\raggedleft}p{80px} >{\raggedleft}p{80px}}
\toprule
\midrule
   & \multicolumn{2}{c}{Default risk, low$^{\text b}$}  & \multicolumn{2}{c}{Default risk, high$^{\text c}$}  \ \tabularnewline
 \cmidrule(r){2-3}\cmidrule(r){4-5}
 Funding$^{\text a}$  & \multicolumn{1}{c}{Long} 	 & \multicolumn{1}{c}{Short} & \multicolumn{1}{c}{Long} 	 & \multicolumn{1}{c}{Short} \tabularnewline
\midrule
\textit{Borrowing rate} $f^+$		&	& &	& \\
$\;\;\;$ 0 bps & 29.33 (0.12) & 	-25.56 (0.22) & 29.65 (0.22) & 	-25.96 (0.41)	\\
$\;\;\;$ 100 bps & 28.70 (0.15)  & 	-28.73 (0.15) & 29.07 (0.22) & 	-29.08 (0.22)		\\
$\;\;\;$ 200 bps & 28.07 (0.22)  & 	-32.14 (0.36) & 28.47 (0.22)  & -32.43 (0.29)	\\
$\;\;\;$ 300 bps & 27.42 (0.30)  & 	-35.93 (0.68) & 27.88 (0.24)  & 	-36.16 (0.61)	\\
$\;\;\;$ 400 bps & 26.75 (0.41)  & 	-39.95 (1.14) & 27.26 (0.27)  & 	-40.10 (1.09)		\\
\midrule
\textit{Lending rate} $f^-$		&	& &	& \\
$\;\;\;$ 0 bps & 25.53 (0.22) & 	-29.36 (0.11) & 25.95 (0.41)  & -29.66 (0.22)		\\
$\;\;\;$ 100 bps & 28.70 (0.15)  & 	-28.73 (0.15) & 29.07 (0.22)  & 	-29.08 (0.22)		\\
$\;\;\;$ 200 bps & 32.14 (0.37)  & 	-28.10 (0.22) & 32.44 (0.29)  & 	-28.49 (0.22)		\\
$\;\;\;$ 300 bps & 35.94 (0.69) & 	-27.45 (0.31) & 36.19 (0.61) & 	-27.89 (0.24)	\\
$\;\;\;$ 400 bps & 39.99 (1.17) & 	-26.77 (0.42) & 40.17 (1.12)  & 	-27.27 (0.27)		\\
\midrule
\bottomrule
\end{tabular}
\hspace{0.1cm}
\\ \begin{minipage}[t]{335px}\footnotesize{Standard errors of the price estimates are given in parentheses.\\
$^{\text a}$ Ceteris paribus changes in one funding rate while keeping the other fixed to 100 bps.\\
$^{\text b}$ Based on the joint default distribution $D_{\text{low}}$ with low dependence.\\
$^{\text c}$ Based on the joint default distribution $D_{\text{high}}$ with high dependence.\\ }
\end{minipage}
\end{table}

\subsection{Non-linearity Valuation Adjustment}

In this last section we introduce a non-linearity valuation adjustment, NVA. The NVA is -- to the best of our knowledge -- proposed here for the first time and is defined by the difference between the correct price $\bar{V}$ and a version of $\bar{V}$ where non-linearities have been approximated away through blunt symmetrization of rates and possibly a change in the close-out convention from replacement close-out to risk free close-out. This entails a degree of double counting (both positive and negative interest). In some situations the positive and negative double counting will offset each other, but in other cases this may not happen. 
Moreover, as pointed out by \cite{BrigoBuescuMorini}, a further source of double counting might be neglecting the first-to-default time in bilateral CVA/DVA pricing. This is done in a number of industry approximations. 

Let $\hat{V}$ be the resulting price of our full pricing algorithm when we replace both $f^+$ and $f^-$ by $\hat{f}$ and we adopt a risk free close-out at default. A further approximation in $\hat{V}$ could be to neglect the first-to-default check in the close-out, along with the above point. We have the following
 \begin{definition}{ \bf (Non-linearity Valuation Adjustment, NVA)}
\[ \mbox{NVA}_t \triangleq \bar{V}_t - \hat{V}_t \]
\end{definition}
As an illustration, we revisit the above example of an equity call option and analyze the NVA in a number of cases. The results are reported in Tables \ref{p4:tab:NVA1}-\ref{p4:tab:NVA2}.
Since we also here adopt a risk free close-out, the examples highlight primarily the double counting error due to symmetrization of borrowing and lending rates. 

\begin{table}
\centering\small
\begin{minipage}[t]{390px}\caption{NVA with default risk and collateralization}\label{p4:tab:NVA1}
\end{minipage}
\begin{tabular}{ccccccc}%p{80px} >{\raggedleft}p{80px} >{\raggedleft}p{80px} >{\raggedleft}p{80px} >{\raggedleft}p{80px}}
\toprule
\midrule
& & & \multicolumn{2}{c}{Default risk, low$^{\text a}$}  & \multicolumn{2}{c}{Default risk, high$^{\text b}$}  \ \tabularnewline
 \cmidrule(r){4-5}\cmidrule(r){6-7}
 \multicolumn{3}{l}{Funding Rates}  & \multicolumn{1}{c}{Long} 	 & \multicolumn{1}{c}{Short} & \multicolumn{1}{c}{Long} 	 & \multicolumn{1}{c}{Short} \tabularnewline
\midrule
$f^+$&$f^-$&$\hat f$		&	& &	& \\
$300$ bps&$100$ bps &$200$ bps & -3.27 (11.9\%) & -3.60 (10.5\%) & -3.16 (11.4\%)   & -3.50 (10.1\%) \\
$100$ bps&$300$ bps &$200$ bps & \,3.63 (10.6\%) & \,3.25 (11.8\%)	& \,3.52 (10.2\%)  & \,3.13	(11.3\%)	\\
\midrule
\bottomrule
\end{tabular}
\hspace{0.1cm}
\\ \begin{minipage}[t]{390px}\footnotesize{The percentage of the total call price corresponding to the NVA is reported in parentheses.\\
$^{\text a}$ Based on the joint default distribution $D_{\text{low}}$ with low dependence.\\
$^{\text b}$ Based on the joint default distribution $D_{\text{high}}$ with high dependence.\\ }
\end{minipage}
\end{table}

\begin{table}
\centering\small
\begin{minipage}[t]{390px}\caption{NVA with default risk, collateralization and rehypothecation}\label{p4:tab:NVA2}
\end{minipage}
\begin{tabular}{ccccccc}%p{80px} >{\raggedleft}p{80px} >{\raggedleft}p{80px} >{\raggedleft}p{80px} >{\raggedleft}p{80px}}
\toprule
\midrule
& & & \multicolumn{2}{c}{Default risk, low$^{\text a}$}  & \multicolumn{2}{c}{Default risk, high$^{\text b}$}  \ \tabularnewline
 \cmidrule(r){4-5}\cmidrule(r){6-7}
 \multicolumn{3}{l}{Funding Rates}  & \multicolumn{1}{c}{Long} 	 & \multicolumn{1}{c}{Short} & \multicolumn{1}{c}{Long} 	 & \multicolumn{1}{c}{Short} \tabularnewline
\midrule
$f^+$&$f^-$&$\hat f$		&	& &	& \\
$300$ bps&$100$ bps &$200$ bps & -4.02 (14.7\%) & -4.45 (12.4\%) & -3.91 (14.0\%)   & -4.35 (12.0\%) \\
$100$ bps&$300$ bps &$200$ bps & \,4.50 (12.5\%) & \,4.03 (14.7\%)	& \,4.40 (12.2\%)  & \,3.92	(14.0\%)	\\
\midrule
\bottomrule
\end{tabular}
\hspace{0.1cm}
\\ \begin{minipage}[t]{390px}\footnotesize{The percentage of the total call price corresponding to the NVA is reported in parentheses.\\
$^{\text a}$ Based on the joint default distribution $D_{\text{low}}$ with low dependence.\\
$^{\text b}$ Based on the joint default distribution $D_{\text{high}}$ with high dependence.\\ }
\end{minipage}
\end{table}

%\begin{table}
%\begin{verbatim}
%Double--Counting Valuation Adjustment	 	 	 
%Low default dependence	High default dependence
% 	Long call	Short call	Long call	Short call
%w/ collateral				
%f+=300,f-=100; f^=200	-3.27	-3.60	-3.16	-3.50
%f+=100,f-=300; f^=200	3.63	3.25	3.52	3.13
%w/ collateral+rehyp				
%f+=300,f-=100; f^=200	-4.02	-4.45	-3.91	-4.35
%f+=100,f-=300; f^=200	4.50	4.03	4.40	3.92
%\end{verbatim}
%\caption{2CVA for...}\label{tab:2cvaexample}
%\end{table}

We can see that, depending on the direction of the symmetrization, NVA may be either positive or negative, and the order of magnitude is about 10-15 percent of the full funding inclusive deal price $\bar V$. This is definitely a relevant figure, especially in a valuation context.

\section{Conclusions: Bilateral Prices or Nonlinear Values?}\label{p4:sec:conclusion}

We have developed an arbitrage-free framework for consistent valuation of derivative trades with collateralization, counterparty credit risk, and funding costs. Based on the risk-neutral pricing principle, we derived a general pricing equation where CVA, DVA, LVA, and FVA are introduced by simply modifying the payout cash-flows of the deal. The pricing problem is non-linear and recursive. The price of the deal depends on the trader's funding strategy, while to determine the funding strategy we need to know the deal price itself. This means that FVA is not generally an additive adjustment, let alone a discounting spread, as commonly assumed by most market participants in simplified approaches. 

Despite starting from a risk neutral formulation, the continuous-time limit of our equations does not depend on unobservable rates, such as the risk-free rate, but only on directly observable rates such as credit spreads, CSA rates, and bank treasury funding rates.
Our valuation framework addresses common market practices of ISDA and CSA governed deals without restrictive assumptions on collateral margin payments and close-out netting rules, and can be also tailored to pricing trades under CCPs with initial and variation margins, as shown in \cite{BrigoPallaviciniJFE}. In particular, we allow for asymmetric collateral and funding rates.

The introduction of funding risk breaks the bilateral nature of the deal price. The particular funding policy chosen by the client is not known to the dealer, and vice versa. As a result, the price of the deal would be different to the two parties. Theoretically, this should mean that the parties would never close the deal, but in reality dealers say this was the key factor driving bid-ask spreads wider during the crisis. 

A further consequence of the non-linearity of the valuation equation is that portfolio asset values do not add up. Valuation is not additive for portfolios in presence of funding costs and/or default risk with replacement close-out. This makes valuation aggregation--dependent and has organizational consequences. Given the impossibility to split valuation adjustments into purely additive credit and funding ones, it is difficult for banks to create CVA and FVA desks with separate and clear-cut responsibilities. This can be done only at the cost of double counting. To initiate an analysis of this problem, we introduced a non-linearity valuation adjustment (NVA) measuring the degree of valuation error one has when removing non-linearities through symmetrization of borrowing and lending rate and through a risk-free close-out substituting a replacement close-out at default.  

We show that the general non-linear pricing equation under credit, collateral and funding can be cast as a set of backward-iterative equations. These equations can be solved recursively by means of least-squares Monte Carlo and we propose such a simulation algorithm. We apply our numerical framework to the benchmark case of the Black-Scholes model for options on equity. The precise patterns of funding adjustments depend on a number of factors, including the asymmetry between borrowing and lending rates. We stress such parameters to analyze their impact on the funding inclusive price. Our numerical results confirm that funding risk does have a non-trivial impact on the deal price and that double counting can be relevant as well.

As a concluding remark, we point out that the violation of the bilateral nature of valuation and aggregation dependence, among other issues, lead to doubts about the funding inclusive value being a price. This prompts to the old distinction between price and value and funding costs could be considered as a profitability/cost analysis tool rather than as a quantity used to charge an adjustment to a client. This is often reinforced by the fact that a bank client has no direct control on the funding policy of a bank, and would therefore have no way to influence or remedy potential inefficiencies for which she would have to pay if charged. 

%\section*{Acknowledgments}
%We are grateful to Claudio Albanese, Marco Bianchetti, Enrico Biffis, Cristin Buescu, Antonio Castagna, Lorenzo Cornalba, St\'ephane C\'repey, Cyril Durand, Nicola Moreni and Giulio Sartorelli for helpful discussions.

%\bibliographystyle{chicago}

\bibliography{arxiv_Funding_paper_online}

\end{document}